\documentclass{article}

\usepackage{caption}
\usepackage{subcaption}
\usepackage{arxiv}
\usepackage{graphicx}
\usepackage[utf8]{inputenc} 
\usepackage[T1]{fontenc}    
\usepackage{hyperref}       
\usepackage{url}            
\usepackage{booktabs}       
\usepackage{amsfonts}       
\usepackage{nicefrac}       
\usepackage{microtype}      
\usepackage{lipsum}
\usepackage[symbol]{footmisc}
\usepackage{amsmath}
\usepackage{amssymb}
\usepackage{algorithm}
\usepackage{arevmath}     
\usepackage{parskip}
\usepackage{amssymb}
\usepackage{setspace}

\title{The efficient market hypothesis for Bitcoin in the context of neural networks}

\author{
Mike Kraehenbuehl \\
  School of Engineering\\
  Zurich University of Applied Sciences\\
  Winterthur, Switzerland \\
  \texttt{kraehmik@students.zhaw.ch} \\
\And
 Joerg Osterrieder \footnotemark[1]\\
  School of Engineering\\
  Zurich University of Applied Sciences\\
  Winterthur, Switzerland \\
  \texttt{joerg.osterrieder@zhaw.ch} \\
}

\begin{document}
\maketitle

\begin{abstract}
This study examines the weak form of the efficient market hypothesis for Bitcoin using a feedforward neural network. Due to the increasing popularity of cryptocurrencies in recent years, the question has arisen, as to whether market inefficiencies could be exploited in Bitcoin. Several studies we refer to here discuss this topic in the context of Bitcoin using either statistical tests or machine learning methods, mostly relying exclusively on data from Bitcoin itself. Results regarding market efficiency vary from study to study. In this study, however, the focus is on applying various asset-related input features in a neural network. The aim is to investigate whether the prediction accuracy improves when adding equity stock indices (S\&P 500, Russell 2000), currencies (EURUSD), 10 Year US Treasury Note Yield as well as Gold\&Silver producers index (XAU), in addition to using Bitcoin returns as input feature. As expected, the results show that more features lead to higher training performance from 54.6\% prediction accuracy with one feature to 61\% with six features. On the test set, we observe that with our neural network methodology, adding additional asset classes, no increase in prediction accuracy is achieved. One feature set is able to partially outperform a buy-and-hold strategy, but the performance drops again as soon as another feature is added. This leads us to the partial conclusion that weak market inefficiencies for Bitcoin cannot be detected using neural networks and the given asset classes as input. 
Therefore, based on this study, we find evidence that the Bitcoin market is efficient in the sense of the efficient market hypothesis during the sample period. We encourage further research in this area, as much depends on the sample period chosen, the input features, the model architecture, and the hyperparameters.

\end{abstract}

\footnotetext[1]{
Financial support by the Swiss National Science Foundation within the project “Mathematics and Fintech - the next revolution in the digital transformation of the Finance industry” is gratefully acknowledged by the corresponding author.
This research has also received funding from the European Union's Horizon 2020 research and innovation program FIN-TECH: A Financial supervision and Technology compliance training programme under the grant agreement No 825215 (Topic: ICT-35-2018, Type of action: CSA) and from Innosuisse under the grant agreement "Strengthening Swiss Financial SMEs through Applicable Reinforcement Learning" (Innosuisse Innovation project 47959.1 IP-SBM).
Furthermore, this article is based upon work from the COST Action 19130 Fintech and Artificial Intelligence in Finance, supported by COST (European Cooperation in Science and Technology), www.cost.eu (Action Chair: Joerg Osterrieder).\newline
The authors are grateful to the management committee members of the COST (Cooperation in Science and Technology), Action Fintech, and Artificial Intelligence in Finance as well as speakers and participants of the $4^{\text{th}}$, $5^{\text{th}}$, and $6^{\text{th}}$ European COST Conference on Artificial Intelligence in Finance and Industry.}

\newpage
\begin{doublespacing}
\tableofcontents
\end{doublespacing}

\newpage
\section{Introduction}
Cryptocurrencies have become popular over the past few years, both among retail and institutional investors. Many studies, but not all, conclude that traditional markets such as stocks or commodities are highly efficient. Therefore it has become tricky to outperform these markets \cite{malkiel_2005}. However, since crypto markets have emerged only in the past ten years, there is a possibility that the markets for this new asset class are not (yet) efficient. \\
In this study, we challenge the weak form of the efficient market hypothesis (EMH) for Bitcoin with the help of a neural network. This has already been tried in a large number of works with different machine learning techniques, various features, and different outcomes \cite{Wildi_2019},  \cite{Huang_2019}, \cite{Mudassir_2020}, \cite{Jaquart_2021}, \cite{Chen_2020}. However, most of the attempts deal with features that are directly derived from the asset under investigation - namely technical features. Less focus is placed on analyzing the connection of other financial instruments with Bitcoin. By examining the following two research questions, we try to contribute to this field of research:

\textbf{RQ 1:} Do additional input features improve the prediction accuracy?

\textbf{RQ 2:} Is the Bitcoin market efficient as per the weak form of the efficient market hypothesis?

First, we give an overview of the relevant literature about the EMH. Second, we review some of the papers that examined Bitcoin and its efficiency. Third, we point out recent papers that used machine learning techniques to predict Bitcoin, as well as other asset classes. Finally, we investigate the research questions with a feedforward neural network with the main focus on analyzing if additional input features improve prediction accuracy. 

\section{Literature Overview}

\subsection{Efficient Markets}

Fama introduces the efficient market hypothesis (EMH) in 1965 \cite{Fama_1965a}, \cite{Fama_1965b}. Another contributor to the EMH is Samuelson \cite{Samuelson_1965}. According to Fama, markets are efficient, only if the prices always “fully reflect” all available information. Further, Fama separates the theory into three parts \cite{Fama_1970}: the weak form (prices reflect all information about past prices), semi-strong (prices reflect all public information: earnings, fundamentals, etc.), and strong form (all private information is reflected in the prices). The following statements can be derived from this:

\begin{itemize}
    \item{Market prices are accurate.}
    \item{A consistent outperformance by using already known information is not possible.}
\end{itemize}

Fama describes the sufficient conditions for a capital market to be efficient as follows; (1) no transaction costs, (2) all information is available to all market participants without any costs, and (3) all market participants agree on the implications of the current information for the current price and the distributions of future prices of each security. Such a market would be efficient. These conditions are sufficient but not necessary for market efficiency. \\
It is also important to note that most economists agree that the EMH is not absolute and has its limitations \cite{EMH_limitations_2015}. Charlie Munger thinks that it is "roughly right" that the stock market is efficient, but not completely efficient \cite{munger1999}. The efficiency in the markets comes from its participants, the investors and traders, who try to make a profit out of every potential information advantage. This causes the prices to include all the information and fully reflect the intrinsic value of its underlying. As markets become more efficient, their returns start to become more and more unpredictable. Grossman and Stiglitz (1980) \cite{Stiglitz_1980} conclude that markets can never be fully efficient. If a market was efficient, no trader would make the effort to gather exclusive information since no outperformance is possible anyway. On the other hand, for a market to be efficient, there have to be enough participants to drive prices to the intrinsic values and consequently make them efficient.
According to Lo (2004) \cite{AMH_2004}, the efficient market hypothesis is incomplete because, during times of market distortions, investors act irrationally which creates inefficiencies that can be exploited by others. To describe this circumstance, he proposes the adaptive market hypothesis (AMH).
On the other hand, Hsieh (2010) \cite{Efficiency_Rationality_2010} claims that market participants may be irrational at some points (e.g. by being systematically biased), but the market can still be efficient, as long as the rational market participants cannot predict the price movements.

\subsection{Bitcoin and Efficiency}

Nakamoto first describes the concept of Bitcoin in his paper in 2008 \cite{nakamoto}. In the following years, Bitcoin becomes more and more popular and scientists start to study this new asset class. Garcia et al. (2014) \cite{garcia_2014} study social feedback related to Bitcoin. Hencic and Gourieroux (2014) \cite{Hencic_2015} predict Bitcoin with a non-causal regressive model. Sapuric and Kokkinaki (2014) \cite{Sapuric_2014} measure the different volatilities of Bitcoin against several major exchange rates of currencies.
Kristoufek (2015) \cite{Kristoufek_2015} studies how the price of Bitcoin is formed and what the drivers of its price are.
Osterrieder and Lorenz (2017) \cite{osterrieder_statistical_2017_lorenz} conclude that the Bitcoin return distribution exhibits higher volatility, stronger non-normal characteristics, and heavier tails than those of the G10 currencies. Chan et al. (2017) \cite{Osterrieder} fit the exchange rates of the largest (by market capitalization) cryptocurrencies against the USD to various parametric distributions. They find that none of the cryptocurrencies analyzed fit a normal distribution and that for Bitcoin the best fit is the hyperbolic distribution. 
Enoksen et al. (2020) \cite{Enoksen_2020} research the emergence of bubbles in the Bitcoin market and other cryptocurrencies.
Most financial assets were already examined for their efficiency. However, these findings may not be valid for the new asset class of cryptocurrencies.
One of the first to investigate whether the Bitcoin price behaves as it would in an efficient market is Urquhart (2016) \cite{Urquhart_2016}. The author conducts various statistical tests like Ljung-Box, AVR test, etc. over the years from 2010 until 2016. He concludes that the Bitcoin market is not weakly efficient during this period. However, it appears that the market becomes less inefficient during the second half of the test window and returns become random.
A study that follows after Urquhart (2016) is "On the inefficiency of Bitcoin" by Nadarajah and Chu (2017) \cite{Nadarajah_2017}. The authors use the same data as Urquhart but continue with a power transformation of the Bitcoin returns. With the transformed data the tests used by the authors show that the Bitcoin market is indeed efficient.
Bariviera (2017) \cite{Bariviera_2017} uses the Hurst exponent to examine daily Bitcoin prices from August 2011 to February 2017. He concludes that the Bitcoin market becomes, at least weak form, efficient after 2014.
Brauneis and Mestel (2018) \cite{Brauneis_2018} investigate the efficiency of a large number of cryptocurrencies. They find that efficiency is positively related to liquidity. Among all cryptocurrencies tested, Bitcoin is the one to pass the most tests of price randomness. Bitcoin is the least predictable and therefore the most efficient.
Tiwari et al. (2018) \cite{Tiwari_2018} find that the Bitcoin market is efficient except for the periods of April–August 2013 and August–November 2016.
Al-Yahyaee et al. (2018) \cite{ALYAHYAEE_2018} show that among stocks, gold, and currencies, Bitcoin is the least efficient.
Sensoy (2018) \cite{Sensoy_2018} also analyzes the weak form of the EMH for Bitcoin and finds that on an intraday level Bitcoin becomes more informationally efficient. Additionally, he states that BTCUSD is slightly more efficient than BTCEUR in the sample period. Two important findings from the study are that the higher the frequency, the lower the pricing efficiency is, and liquidity (volatility) has a significantly positive (negative) effect on the efficiency of Bitcoin prices.
Studies like \cite{Urquhart_2016} conclude that Bitcoin becomes more efficient over time and also the correlation with stock indices increases. This pattern is not surprising as young markets, like some emerging markets, tend to become more efficient over time as more investors and traders enter the market and price in the information that they believe to have. 
Vidal-Tomás and Ibañez (2018) \cite{vidal_2018} study the semi-strong form of efficiency. They find that Bitcoin does not respond to measures of central banks, but becomes increasingly efficient in regards to negative and positive news related to Bitcoin itself.
Khuntia and Pattanayak (2018) \cite{Khuntia_2018} test the adaptive market hypothesis on the Bitcoin market where they verify the evolving efficiency in the Bitcoin market. Major events coincide with times of efficiency and inefficiency, therefore the AMH rather than the EMH seems to apply to the behavior of the Bitcoin price.
Bitcoin is probably also less efficient/rational than the stock and bond markets because of the lack of intrinsic value. Investors can derive a fair price for a stock at least in part from the company's key figures, revenues, profits, etc. This is not possible for cryptocurrencies to the same extent, and therefore much more speculation is involved. However, this can be a sign of irrationality, but not necessarily of inefficiency.

All in all, there is mixed evidence for the predictability of Bitcoin. It seems as if the Bitcoin market is becoming more efficient as it matures.

\subsection{Machine Learning in Financial Markets}

Machine learning (ML) methods enjoy increasing popularity in applications used for empirical asset pricing due to their ability to learn complex relationships between features and target. Gu et al. (2020) \cite{Shihao_2020} identify neural networks and regression trees are the best performing methods due to their advantage of accommodating nonlinear interactions in contrast to other ML methods. As the most powerful predictors for stock prices, the authors mention the ones showing price trends including return reversal and momentum, liquidity, volatility, and valuation ratios.
Wildi and Bundi (2019) \cite{Wildi_2019} analyze the inefficiency of Bitcoin using several methods including ARMA and neural networks. They use a feedforward neural net, with Bitcoin lags 1-6 as inputs. The architecture with only two hidden layers containing 6 and 3 nodes respectively is rather simple. However, they are still able to generate an outperformance when looking at the ensemble average of random nets. Overall they strongly reject the EMH for Bitcoin. In contrast to other before-mentioned studies, they observe increasing market inefficiency towards the end of the sample period which is 2019-01-10.
Huang et al. (2019) \cite{Huang_2019} use Bitcoin price-based high dimensional technical indicators in their research. With a classification tree-based model and 124 technical indicators, they provide evidence for strong out-of-sample performance for narrow ranges of daily Bitcoin returns. They find that technical analysis can be useful when predicting returns for an asset like Bitcoin that is hard to value by fundamentals.
Mudassir et al. (2020) \cite{Mudassir_2020} use the four models ANN, SANN, SVM, and LSTM to research different prediction horizons for Bitcoin. The classification models for next day forecasts score up to 65\% accuracy. Jaquart et al. (2021) \cite{Jaquart_2021}, however, criticize the authors for creating a potentially unbalanced training dataset. For financial time series, which are usually noisy, an unbalanced dataset can lead to a majority prediction of a class that is overrepresented in the training dataset. If the same class is also overrepresented in the test dataset, this can automatically lead to higher accuracy.
Jaquart et al. express similar critics to Chen et al. (2020) \cite{Chen_2020} that find that simpler methods like logistic regression perform better in the 1-day forecast than more complex ones like recurrent neural networks.
In their work, Jaquart et al. (2021) \cite{Jaquart_2021} analyze the intraday Bitcoin price on prediction horizons from 1 to 60 minutes. Their approach is to use different machine learning models like recurrent networks and gradient boosting among others. Moreover, they use four different feature sets; technical features derived from historical Bitcoin data, blockchain-related features, sentiment-/interest-based features, and asset-based features other than Bitcoin (e.g. gold and MSCI World index returns). Their findings are that technical features are most important for most methods. An outperformance against a random classifier is achieved but this turns into negative returns after accounting for transaction costs. Therefore, this study does not violate the efficient market hypothesis.
Feng and Giglio (2020) \cite{Guanhao_2020} introduce a methodology to systematically evaluate the marginal contribution of a new factor relative to the existing factors in explaining asset prices. They use techniques like the double-selection LASSO method \cite{Belloni} and two-pass regressions such as Fama-MacBeth for their procedure. The proposed method can be used to identify a good set of potentially market-predictive features.

Comparing performances of predictive models from different works is often difficult. The reason for this is that one can use different time horizons, target and explanatory variables, parameter specifications, and evaluation metrics.

\section{Data and Summary Statistics}

This section will discuss the input data that is used for the experiment in section \ref{sec:experiment}. With an artificial neural network, we use different feature sets to predict the Bitcoin price movement for the next day. The predictive performances of the feature sets are subsequently compared.

\subsection{Features}

There are several studies that use technical features derived from its underlying to predict Bitcoin, and they show some evidence for their predictive power. \cite{Jaquart_2021}, \cite{Chen_2020} use other asset-based features such as major stock indices, interest rates, or currency market data. With our study, we would like to contribute to research in this direction and therefore investigate the six features in Table \ref{table:1} to validate the two research questions stated at the beginning. 


The daily prices of selected features are imported from Yahoo Finance (YF) with a starting date of 2014-09-17. This period should be sufficient to answer the research questions because, as mentioned in the previous section, some studies conclude that Bitcoin was not efficient in the early years and has become more efficient over time. 

\begin{table}[ht]
\centering
\begin{tabular}{ p{4.3cm}p{2.9cm}p{2cm}p{2.3cm}}
Feature & Source & YF Symbol & Timestamp \\
\hline
BTC/USD & CoinMarketCap & BTC-USD & 23:59 UTC \cite{coinbase_closing}  \\
S\&P 500 & Standard \& Poors & \^{}GSPC & 20:00 UTC \cite{NYSE_closing}  \\
Russell 2000 & FTSE Russell&\^{}RUT & 20:00 UTC \cite{NYSE_closing} \\
EUR/USD & ICE Data Services &EURUSD=X & 21:30 UTC \cite{EURUSD_YF}\\
10Y Treasury Note Yield Index & CBOE& \^{}TNX & 19:00 UTC \cite{Treasury_YF} \\
PHLX Gold\&Silver Index& NASDAQ & \^{}XAU& 21:30 UTC \cite{xau_bloomberg}\\
\hline
\end{tabular}
\caption{Selected input features and source.}
\label{table:1}
\end{table}%

Bitcoin is the largest cryptocurrency by market cap and accounts for around 45\% of the total USD 1.28 trillion in the crypto market \cite{Coinbase_Marketshare}. On the one hand,  we will use Bitcoin as a target for our predictive model, and on the other hand, we will also use past Bitcoin returns as input features. CoinMarketCap reports the Bitcoin price (BTC/USD) by calculating a volume-weighted average of all market pair prices reported for Bitcoin \cite{Coinbase_prices}. The prices are reported from crypto exchanges like Binance, Coinbase, and FTX among others. 

The S\&P 500 Index represents the market price of the 500 biggest (by market cap) companies in the United States. Its value is reported whenever the NYSE is open for trading. The closing price of the S\&P 500 Index is calculated according to the closing prices and the weightings of the included stocks. Kim et al. (2020) \cite{Kim_2020} show in an empirical study that there exists a time-varying relationship between the Bitcoin market and the S\&P 500. Therefore the S\&P 500 price may also provide information to predict future Bitcoin price movements. 

Fama and French (1993) \cite{FAMA_Riskfactors_1993} outline several risk factors in the returns of stocks and bonds. One of the factors is the size, or that small companies outperform large ones. Therefore, the Russell 2000 Index will be used in the experiment to account for this risk factor. The Russell 2000 Index is a subset of the 2000 smallest companies that are represented in the Russell 3000 Index \cite{FTSE_Russell}. The companies represented by the Russell 2000 Index only account for 10\% of the total market capitalization of the Russell 3000 Index. Its closing price is calculated daily based on the closing prices at the NYSE of the stocks contained in the index.

Currencies may also add value to the prediction of cryptocurrencies. The EUR/USD exchange rate is among the biggest and most liquid currency pairs.

The CBOE 10 Year Treasury Note Yield Index (TNX) is based on the yield-to-maturity of the most recently auctioned 10 Year Treasury Note. 

In the before-mentioned study by Kim et al. (2020) \cite{Kim_2020}, they not only show a relationship between Bitcoin and the US stock market but also with gold. For reproducibility reasons, we use the PHLX Gold and Silver Index (XAU). We calculate a 0.95 Pearson correlation with the CME Globex Gold Future on 100 troy ounces \cite{Gold_CME} (YF-symbol: GC=F). If gold and silver prices share a dependency with Bitcoin, the XAU index should also serve as a good input feature for predicting Bitcoin. The Philadelphia Stock Exchange Gold and Silver Index is a capitalization-weighted index that includes the leading companies involved in the mining of gold and silver \cite{xau_bloomberg}. All securities in the index must be listed on The Nasdaq Stock Market, New York Stock Exchange, NYSE American, or Cboe BZX Exchange \cite{XAU_methodology}, \cite{Nasdaq_methodology}.

Furthermore, we analyze the log returns of each input feature as seen in equation \ref{eqn:logreturns}. The main reason for using log returns is that they are stationary and additive. The neural network is then trained with normalized log returns. To describe the distributions of the features, the skewness and kurtosis of each features log returns are calculated in Table \ref{table:2}.

\begin{equation}
    \label{eqn:logreturns}
    \log(1 + r_{t}) = \log\Bigg(\frac{P_t}{P_{t-1}}\Bigg) = \log(P_{t}) - \log(P_{t-1})
\end{equation}

\begin{table}[ht]
\centering
\begin{tabular}{p{3.55cm}p{1.5cm}p{1.5cm}p{1.5cm}p{1.5cm}p{1.5cm}}
Feature &mean&  min &max &skewness &kurtosis \\
\hline
BTC/USD & 0.0015 & -0.465 & 0.225  & -0.767  & 11.112  \\
S\&P 500 & 0.0005 & -0.099 & 0.089 & -0.217 & 13.434 \\
Russell 2000 & 0.0003&-0.118 &0.089& -0.696 & 9.545 \\
EUR/USD & 0.0000 &-0.028 & 0.028& 0.069 &3.236 \\
10Y Treasury Yield & 0.0003& -0.271 & 0.404&1.564& 28.149  \\
Gold/Silver & 0.0002 & -0.146 & 0.144 &-0.059 & 3.314  \\
\hline
\end{tabular}
\caption{Stats of log returns.}
\label{table:2}
\end{table}%

\begin{figure}[ht]
    \centering
    \includegraphics[width=9cm]{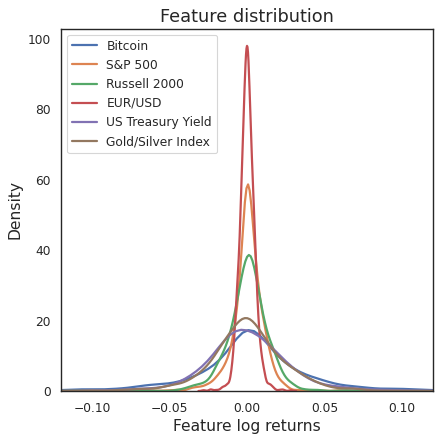}
    \caption{Density plot of the log returns of the features.}
    \label{fig:distributions}
\end{figure}

Bitcoin is notorious for its high volatility compared to other assets. Among the selected features, Bitcoin had the highest daily loss of -0.465. The Fisher kurtosis of 11.112 also shows that the tails are thicker than in the normal distribution (normal distribution has a kurtosis of 0.00). Bitcoin, S\&P 500, Russel 2000, and the Gold/Silver index have negative skewness. On the other hand, the 10Y Treasury Yield has a high positive skewness of 1.564 and the highest kurtosis of 28.149. Between 2020-03-06 and 2020-03-10, both the highest negative log return of -0.271 and the highest positive log return of 0.404 occurred. The S\&P 500 and Russell 2000 distributions are very similar, which is not surprising as both indices represent US companies. Among the assets examined, the EUR/USD and Gold/Silver returns most closely resemble a normal distribution. Fig. \ref{fig:distributions} shows the distributions of the log returns.

\begin{figure}[h]
    \centering
    \includegraphics[width=12cm]{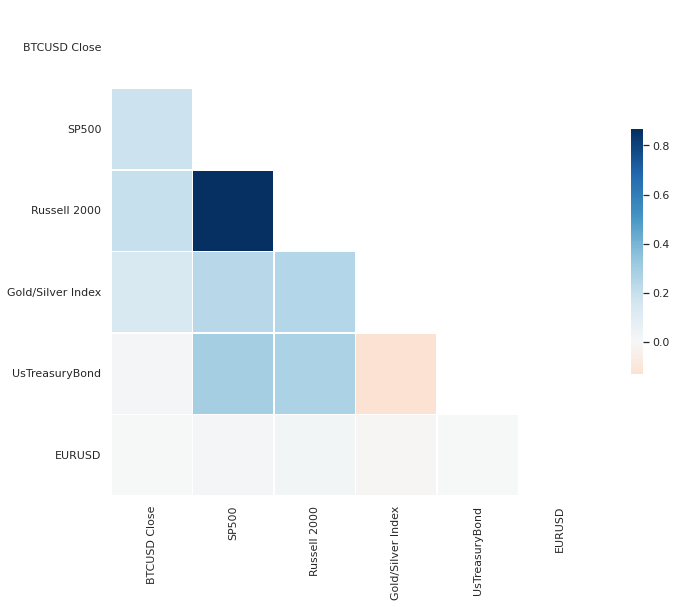}
    \caption{Correlation heatmap}
    \label{fig:heatmap}
\end{figure}

The correlation heatmap in Fig. \ref{fig:heatmap} (appendix) shows a strong correlation between the Russell 2000 and S\&P 500 Index since both indices represent US companies. Other combinations show a moderate positive correlation with two exceptions. First, the 10Y Treasury Yield appears to not correlate with Bitcoin. Second, we can see a negative correlation of Gold/Silver with the 10Y Treasury Yield. This plot only shows a simple correlation and not a possible time-varying nonlinear dependence between the different features. In the appendix \ref{sec:autocorrelation}, we further discuss Bitcoin's autocorrelation and a reason why we do not consider it in this study.
\newpage

\subsection{Missing Values}

All six input features described above are combined into one dataset. The time series range from 2014-09-17 until 2022-04-30. At this point, the dataset has missing values for all features except Bitcoin, which is traded every day without exception. All other assets have no prices published on weekends and public holidays. Additionally, because we use log returns as inputs for the model in the experiment, we produce missing values on Mondays and all other dates with no value for the previous day because no one-day log returns can be calculated.
There are multiple ways how to deal with missing values in machine learning. One method is to replace them with the last known value but this will distort the data. Another common approach is to interpolate to estimate the missing value. In practice, this is not very feasible because only the past values are known and this would also falsify the data. Since a neural network cannot be trained with missing values, the most reasonable consequence, therefore, is to delete all rows in the dataset, containing assets with missing values. After doing so, the initial size of the dataset shrinks from 2783 rows to 1495 rows. 
\\Furthermore, because we want to predict the next day's price movement of Bitcoin, the input dataset with all selected features contains data delayed by one day (lag 1). The problem of missing values intensifies when lag 2 is added to the input dataset. For illustration, if we take the target (Bitcoin signal) for 2022-03-23 (Wednesday) and choose the S\&P 500 as the input feature, we have a value for 2022-03-22 (lag 1) but no value for 2022-03-21 (lag 2). To explain the current Bitcoin price movement, there has to be a lag 1 and lag 2 value for every input feature. If we create a feature dataset that besides lag 1 also contains lag 2, the dataset will shrink to 1092 rows.
With each additional lag added for features other than Bitcoin, more rows contain missing values, therefore more rows have to be deleted which makes the already sparse daily data even smaller. Therefore we stay with lag 1 for this experiment. 
This is unfavorable for training but at least training and predictions are based upon real values. Estimating missing data would distort our analysis in particular the comparison between different feature sets. Using intraday data would be an interesting option as more data points are available and the problem becomes less extensive.

\newpage
\section{Experiment Approach}  \label{sec:experiment}

With an artificial neural network, the next day's Bitcoin price movements (signals) shall be predicted. Neural networks have the advantage that they can deal with nonlinear dependencies. With this experiment, we want to analyze the following research questions:

\textbf{RC 1:} Do additional input features lead to a better result? This will be measured by comparing the effects of different feature sets on prediction accuracy.

\textbf{RC 2:} Is the Bitcoin market efficient under the weak form of the EMH? If the correctly predicted price movements are statistically significant, this will be a sign that the Bitcoin market is not efficient. 

The aim of this experiment is not to parameterize the model in such a way as to generate the best possible outperformance but to examine these two questions under the chosen scenario.

\textbf{Experiment process}

\begin{enumerate}
\item Import daily prices of selected features from Yahoo Finance
\item Calculate log returns
\item Create six different sets of input features
\item Shift (lag) features by one day
\item Drop all rows with missing values
\item Normalize lagged features to a range between 0-1
\item Create binary Bitcoin signals as the target variable
\item Split data into training-, validation-, and test-set
\item Train feedforward neural network with training set to predict next days Bitcoin signal
\item Evaluate model out-of-sample with testing set
\end{enumerate}

\subsection{Feedforward Neural Networks}

\begin{figure}[ht]
    \centering
    \includegraphics[width=12cm]{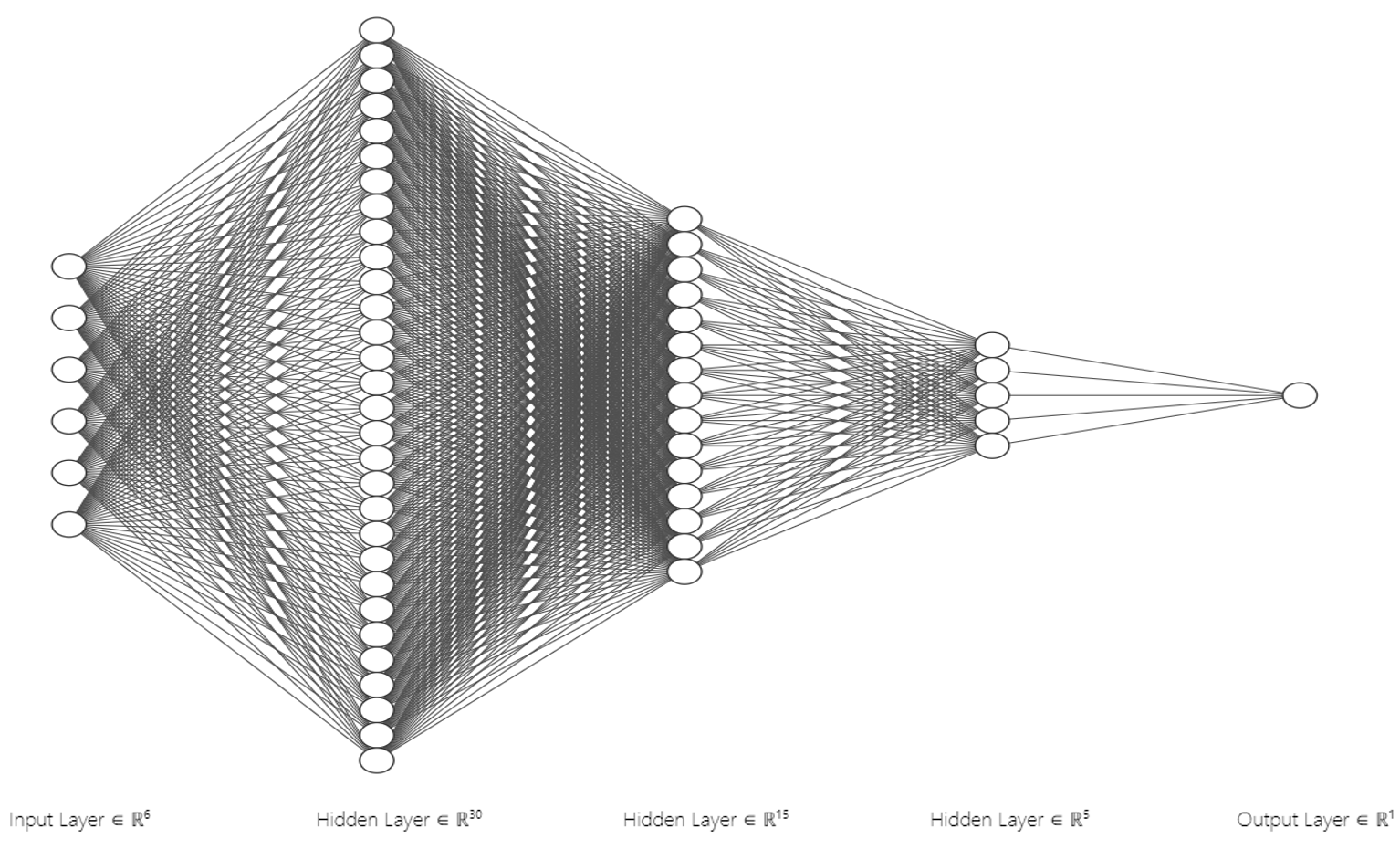}
    \caption{Architecture used for the model \cite{NN_architecture}.}
    \label{fig:NNarchitecture}
\end{figure}

A neural network is designed to mirror the functioning of the human brain. It contains many nodes that are interconnected in a specific way. The individual neurons receive information (input) from other neurons, transform the information and then pass it on (output) to other connected neurons. The basic idea is that the network attempts to construct a function f(x) that maps a given input X to a given target value y. This is is an example of supervised learning, but there are also neural networks that can be used for unsupervised learning \cite{Japkowicz_2001}.
An ever-increasing number of types of neural networks exist with each having its advantages; e.g. perceptron, multilayer perceptron (MLP), feedforward neural network (FNN), convolutional neural networks (CNN),  recurrent neural networks (RNN), long short-term memory (LSTM) \cite{LSTM_1997} and many more. For this experiment, we chose the rather basic feedforward neural network since the focus is not on optimizing to the best possible performance, but rather on comparing different feature sets. Fig. \ref{fig:NNarchitecture} shows the information flow in an FNN. In contrast to recurrent neural networks, feedforward networks only pass information in one direction (forward). The input layer represents the explanatory variables or features (x) that are entered into the system. After that comes any number of hidden layers with each containing any number of neurons. These layers are responsible for processing, transforming, and connecting information. Finally, there is an output layer that generates the predicted y. For the network to differ from a normal linear regression, at least one layer must have an activation function g(x). This function must be either nonlinear or at least only partially linear function, but never completely linear. Equation \ref{eqn:neuralnet} describes the transformation of information from layer (n-1) to the next layer n \cite{Jaquart_2021}.

\begin{equation}
\label{eqn:neuralnet}
    a^{(n)} = g^{(n)}\bigg(W^{(n)}a^{(n-1)}+b^{(n)}\bigg)
\end{equation}

where:

\(a^{(n)}\) is the output of layer n 

\(g^{(n)}\) is the activation function

\(W^{(n)}\) is the weight matrix that weight the values \(a^{(n-1)}\) that have been forwarded from layer n-1 to layer n

\(b^{(n)}\) is the bias for layer n 

\newpage

\subsection{Data Preprocessing}

\textbf{Features}

Data is imported with the Python package yfinance from Yahoo Finance so that the code for the neural network can easily be executed by any user without the need for external datasets. The used data starts on 2014-09-17 and ends on 2022-04-30. The imported dataset of Bitcoin and the selected features consist of prices. At this stage, rows that include N/A values are kept in the dataset. Note that Bitcoin is traded every day. Stocks and currencies on the other hand have missing values at weekends and public holidays. 
To answer the question of additional input features leading to higher prediction accuracy, the sets of input features outlined in Table \ref{table:3} are created. Each feature set contains the log returns of the input features shifted by one day (lag 1) and all rows with missing values are deleted. For the experiment, each feature set will be trained separately with the neural network, and predictions are made individually. Then the results metrics from the different feature sets will be compared. We decide to only use lag 1 for two reasons. Firstly, if more lags are added, the dataset is reduced due to more missing values in the rows. For example, if lag 2 is added, the dataset is reduced by 27\% from 1495 rows to 1092 rows. Since the return from the previous day will most likely be most important, we only choose lag 1 as the input feature. The reason for this decision is described in more detail in the appendix section \ref{sec:autocorrelation}.

\begin{table}[ht]
\centering
\begin{tabular}{p{2cm}p{12cm}}
Feature set&Included features (lag 1)\\
\hline
Feature set 1	&Bitcoin \\
Feature set 2	&Bitcoin, S\&P 500\\
Feature set 3	&Bitcoin, S\&P 500, Russell 2000 \\
Feature set 4	&Bitcoin, S\&P 500, Russell 2000, EUR/USD \\
Feature set 5	&Bitcoin, S\&P 500, Russell 2000, EUR/USD, US Treasury Yield \\
Feature set 6	&Bitcoin, S\&P 500, Russell 2000, EUR/USD, US Treasury Yield, Gold+Silver \\
\hline
\end{tabular}
\caption{Feature sets containing log returns shifted by 1 day (lag 1).}
\label{table:3}
\end{table}%

The feature sets now consist of shifted log returns without missing values. It is common practice to normalize the input features for the training of neural networks. For this study, the log returns of the features are normalized to a range between 0-1 with the MinMaxScaler \cite{MinMaxScaler} equation \ref{eqn:minmaxscaler}. The distribution of the data remains the same.

\begin{equation}
\label{eqn:minmaxscaler}
    X_{sc} = \frac{X - X_{min}} {X_{max} - X_{min}}
\end{equation}

\textbf{Target - binary Bitcoin signal}

The goal of the neural network in this experiment is to predict the Bitcoin price movement of the next day. This will be defined as a binary classification problem in equation \ref{eqn:signal}, where 0 represents a negative Bitcoin return and 1 represents a positive Bitcoin return. For further analysis or in practice applications, 0 can stand for sell and 1 for buy. Or short or long. It would have to be decided what it means when implementing a trading strategy. In this experiment, we use a long-only approach where the threshold between a positive and a negative signal is a probability of 0.5. One could also define that a long position is taken if the probability of a rising price is above 65\%, and a short position is taken if the probability is less than 35\%. In this study, this is ignored when analyzing the effects of the features on accuracy.

\begin{equation}
\label{eqn:signal}
  y_{t} =
    \begin{cases}
      1 & \text{, if return} \ \geq 0\\
      0 & \text{, if return} \ < 0\\
    \end{cases}       
\end{equation}

where \(y_{t}\) is the Bitcoin signal\\

\textbf{Train- validation- and test-set}

\begin{figure}[ht]
    \centering
    \includegraphics[width=11cm]{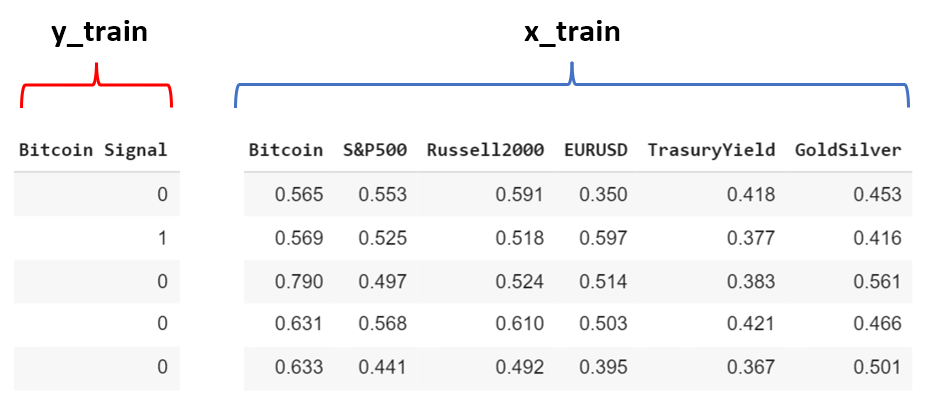}
    \caption{Example of the dataset with binary target and normalized explanatory variables used for training.}
    \label{fig:xytrain}
\end{figure}

Now that the different feature sets are defined, the data is split into separate datasets to meet the requirements of the neural network algorithm. In this case, a train level of 60\% is applied, which means that the first 60\% of data is used for training. For this purpose, the two datasets x\_train and y\_train are generated, which contain the normalized log returns of the selected input features shifted by one day and the binary target variable, respectively (Fig. \ref{fig:xytrain}). The remaining 40\% are used for testing the out-of-sample performance. The test dataset is named x\_test and y\_test. 
Additionally, we define a validation split of 10\%. Therefore, 10\% of the training data is used as a validation set for hyperparameter tuning during the training process. The distribution of values may differ between train, validation, and test dataset. Since in practice this will most likely also be the case we do not see this as a major problem but it is discussed further with the results in section \ref{sec:AccuracyLoss}.

\subsection{Neural Network Architecture}

There exist multiple different types of neural networks, each having its advantages in certain application areas. For time series the most suitable types are feedforward, recurrent, and LSTM. Because the aim of this experiment is not to generate the best possible performance on the sample we use a simple feedforward neural network instead of more complicated models. Wildi and Bundi \cite{Wildi_2019} use a feedforward net with two hidden layers with dimensions six and three. Even with such a simple architecture, the neural network is able to outperform a buy-and-hold strategy, an EqMa(6), and an MA(6) model. The following neural network, accessible on GitHub \cite{Github_kraehenbuehl}, is programmed in Python by using the Keras API \cite{keras}. The neural network is set up with the architecture in Table \ref{table:4}.

\begin{table}[ht]
\centering
\begin{tabular}{p{3cm}p{10cm}}
Layer&Neurons \\
\hline
Input Layer &        with dimension according to the amount of input features\\
First hidden Layer &  30 Neurons with ReLu    activation function\\
Second hidden Layer&  15 Neurons with ReLu    activation function\\
Third hidden Layer &   5 Neurons with ReLu    activation function\\
Output Layer      &    1 Neuron  with Sigmoid activation function\\
\hline
\end{tabular}
\caption{Layers, neurons, and activation functions used for the experiment.}
\label{table:4}
\end{table}%

\newpage

\textbf{ReLU \& Sigmoid activation functions}

\begin{figure}[ht]
  \centering
  \begin{minipage}[b]{0.46\textwidth}
    \includegraphics[width=7.0cm]{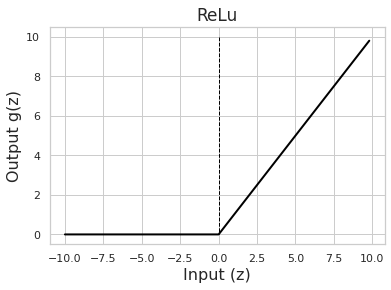}
    \caption{ReLU activation function used in hidden layers.}
    \label{fig:relu}
  \end{minipage}
  \hfill
  \begin{minipage}[b]{0.47\textwidth}
    \includegraphics[width=7.0cm]{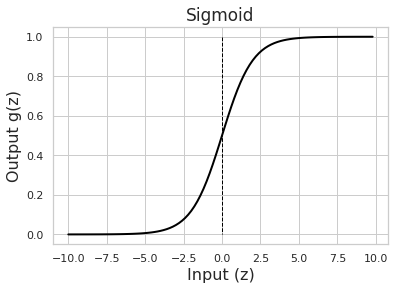}
    \caption{Sigmoid activation function used in the output layer.}
    \label{fig:sigmoid}
  \end{minipage}
\end{figure}

The Rectified Linear Units (ReLU) function \ref{eqn:relu} visualized in Fig. \ref{fig:relu} is widely used in deep learning applications. Despite its simplicity, the function can account for interactions and nonlinearities. Traditional activation functions like tanh and sigmoid can lead to the problem of a vanishing gradient if they are used in many hidden layers. Because their derivatives are very small at most points of the curve, it makes it difficult to change the weights when applying gradient descent. ReLU helps reduce this problem since its derivative is either 0 or 1. This is why ReLU is commonly used as an activation for hidden layers \cite{relu}.

\begin{equation}
\label{eqn:relu}
    \text{ReLU}(z) = \text{max}(0, z)
\end{equation}

The sigmoid function \ref{eqn:sigmoid} visualized in Fig. \ref{fig:sigmoid} maps any real number to a value between 0 and 1. One could also say that it is a function that maps a weighted sum to probabilities. As the sigmoid function is used in the output layer, the output of the sigmoid function is between 0 and 1 and the threshold function \ref{eqn:threshold} maps the value $\sigma(z)$ to 1 if the predicted probability for the Bitcoin signal being positive is greater than 0.5, and vice verca for a negative signal. In this case the prediction 0 means "not-in-the-market" and 1 means "in-the-market".

\begin{equation}
\label{eqn:sigmoid}
    \sigma(z) = \frac{1} {1 + e^{-z}}
\end{equation}

\begin{equation}
\label{eqn:threshold}
  t(p) =
    \begin{cases}
      1 & p \ \geq 0.5\\
      0 & p \ < 0.5\\
    \end{cases}       
\end{equation}

where:

\(z\) is the weighted sum from equation \ref{eqn:neuralnet}\\
\(p\) is the probability output from the sigmoid function \ref{eqn:sigmoid}\\

\textbf{Loss function - Binary cross-entropy}

Different loss functions can be used for training a neural network. The purpose of the loss function is to calculate the error of a given state. A common loss function used for regression is the mean squared error (MSE). The binary cross-entropy (equation \ref{eqn:crossentropy}), also called log-loss, is another loss function, which is designed to measure the performance of a binary classification problem. This function is applied in the output layer after the sigmoid but before the threshold function. The input for the cross-entropy loss function is a value between 0 and 1, or a probability value for the signal being positive. Since signals (0 and 1) have to be predicted in this experiment, the cross-entropy loss function is a better choice than the MSE. 

\begin{figure}[ht]
  \centering
  \begin{minipage}[b]{0.47\textwidth}
    \includegraphics[width=5.8cm]{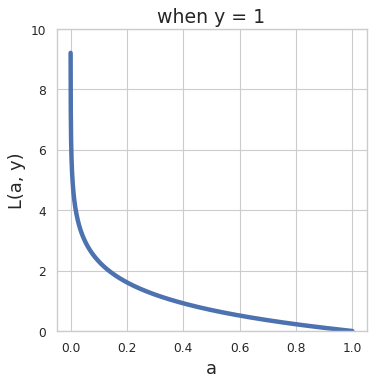}
    \caption{Cross-entropy when target signal is 1.}
    \label{fig:crossentropy1}
  \end{minipage}
  \hfill
  \begin{minipage}[b]{0.47\textwidth}
    \includegraphics[width=5.8cm]{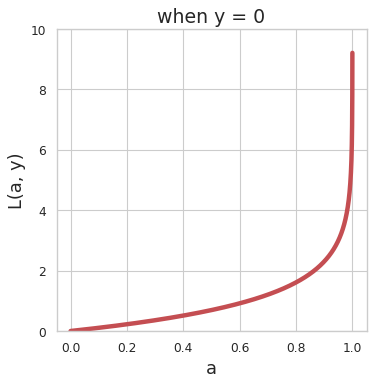}
    \caption{Cross-entropy when target signal is 0.}
    \label{fig:crossentropy0}
  \end{minipage}
\end{figure}

Example: 
If the target signal is 1 (positive log return of Bitcoin), the cross-entropy loss function illustrated in Fig. \ref{fig:crossentropy1} calculates -log(a) as loss value. If the output of the sigmoid function is near 0, the loss will be large. This means that a wrong prediction that is near 0 instead of 1, will be punished strongly. On the other hand, it will drop drastically if it is close to 1. This way, wrong predictions cause a high loss and will therefore update the weights strongly. A small loss will update the weights not much. Fig. \ref{fig:crossentropy0} shows the curve when the target is 0.

\begin{equation}
\label{eqn:crossentropy}
    \mathcal{L}(a, y) = -\Big[y\log(a) + (1 - y)\log(1 - a)\Big]
\end{equation}

where: 

\(y\) is the binary target class (0 or 1)\\
\(a\) is the predicted probability that z belongs to class 1 (equation \ref{eqn:sigmoid})\\

\textbf{Optimizers and hyperparameters}

As large as the selection of network types and architectures is, there is also an enormous choice of hyperparameters and optimizers. ADAM and BatchNormalisation is used in the model. Further, the model is trained with 300 epochs and a batch size of 20 at a learning rate of 0.00005.

\section{Results}

\textbf{Influence of randomness}

The neural network's algorithm starts with initial weights that are adjusted during training according to the training data and the defined network specifications. Depending on how these weights are initiated at the beginning, the outcome of the final weights and therefore the predictions and performance on the test dataset will be different. Therefore, one could just run the model many times until a good set of initial weights was initiated so that the performance on the test set will also be good. This would distort the meaning of out-of-sample testing. Therefore, for this experiment, each feature set is trained and tested 50 times. For later result comparison of the feature sets, we then take averages of the metrics of the 50 iterations. This is how we try to make a representative comparison.
To show the importance of randomness, we take a look at the best (iteration 36/50) and worst (iteration 40/50) model outcomes of feature set 2 (Bitcoin lag 1 + S\&P 500 lag 1). The data and model settings have been the same, except for the different initial weights. The best iteration achieves an out-of-sample accuracy of 54.5\% and a logarithmic cumulative sum of 1.52 while the worst iteration only has an accuracy of 49\% and a logarithmic cumulative sum of 0.13. 

\textbf{Test model quality on same day data}

The model is tested on same-day Ethereum data to prove that the model would generate reasonable predictions with inputs that actually have a dependency on the target. Ethereum is the second-largest cryptocurrency and is highly correlated with Bitcoin. We create a set of training features that only consists of Ethereum log returns with the same timestamp as Bitcoin (same day). The data is processed and used in the model as described above. We refer to the appendix section \ref{sec:ETH} for the results of the same-day Ethereum feature set. Tested 10 times, the average prediction accuracy for the Bitcoin signal is 80.5\%. Therefore, we assume that the model would recognize existing dependencies between input features and target variable (if there are any) and generate prediction accuracies higher than a buy-and-hold strategy.

\textbf{Results of 50 iterations}

The feature set 0 "rNorm" is created to serve as a benchmark. It consists of a sample of randomly drawn normally distributed values where the mean is 0 and the variance is equal to the variance of the Bitcoin log returns. Subsequently, the feature set rNorm is processed and evaluated in the same way as the other feature sets. Each feature set in Table \ref{table:5} is trained and tested 50 times and the averages of the resulting metrics are saved in Table \ref{table:6}. The plots of the individual feature sets with their results can be found in the appendix section \ref{sec:results}.

\begin{table}[ht]
\centering
\begin{tabular}{p{2cm}p{13.3cm}}
Feature set&Included features (lag 1)\\
\hline
Feature set 0	&$\text{rNorm} \sim \mathcal{N}(\mu, \,\sigma^{2})$  \hspace{0.2cm} , where $\mu=0$ and $\sigma^{2}=\text{var(Bitcoin log returns)}$\\
Feature set 1	&Bitcoin \\
Feature set 2	&Bitcoin, S\&P 500\\
Feature set 3	&Bitcoin, S\&P 500, Russell 2000 \\
Feature set 4	&Bitcoin, S\&P 500, Russell 2000, EUR/USD \\
Feature set 5	&Bitcoin, S\&P 500, Russell 2000, EUR/USD, US Treasury Yield \\
Feature set 6	&Bitcoin, S\&P 500, Russell 2000, EUR/USD, US Treasury Yield, Gold+Silver \\
\hline
\end{tabular}
\caption{Feature sets 1-6 containing log returns shifted by 1 day (lag 1).}
\label{table:5}
\end{table}%

\begin{table}[ht]
\centering
\begin{tabular}{p{1.9cm}p{1.6cm}p{1.6cm}p{1.6cm}p{1.6cm}p{1.5cm}p{1.5cm}p{1.6cm}}
Feature set&Accuracy (in-sample)&Accuracy (out-of-sample)& Loss \hspace{0.4cm} (in-sample)&Loss \hspace{0.4cm}(out-of-sample)  & Log-\hspace{0.2cm} cumsum&  Sharpe\hspace{0.2cm} ratio &Pos. / neg. signals forecasted \\
\hline
Feature set 0 & 0.549 & 0.525 & 0.687 & 0.698 & 0.922 & 0.603 & 534 / 64 \\
Feature set 1 & 0.546 & 0.527 & 0.693 & 0.701 & 0.938 & 0.687 & 499 / 99 \\
Feature set 2 & 0.574 & 0.519 & 0.678 & 0.712 & 0.870 & 0.696 & 472 / 126 \\
Feature set 3 & 0.585 & 0.523 & 0.671 & 0.720 & 0.812 & 0.646 & 453 / 145 \\
Feature set 4 & 0.595 & 0.530 & 0.663 & 0.719 & 1.109 & 0.866 & 424 / 174 \\
Feature set 5 & 0.603 & 0.527 & 0.659 & 0.730 & 0.833 & 0.637 & 400 / 198 \\
Feature set 6 & 0.610 & 0.525 & 0.654 & 0.733 & 0.853 & 0.656 & 392 / 206 \\
\hline
\end{tabular}
\caption{Results averaged over 50 iterations per feature set.}
\label{table:6}
\end{table}%

\subsection{Accuracy \& Loss} \label{sec:AccuracyLoss}
\begin{figure}[ht]
  \centering
  \begin{minipage}[b]{0.43\textwidth}
    \includegraphics[width=6.8cm]{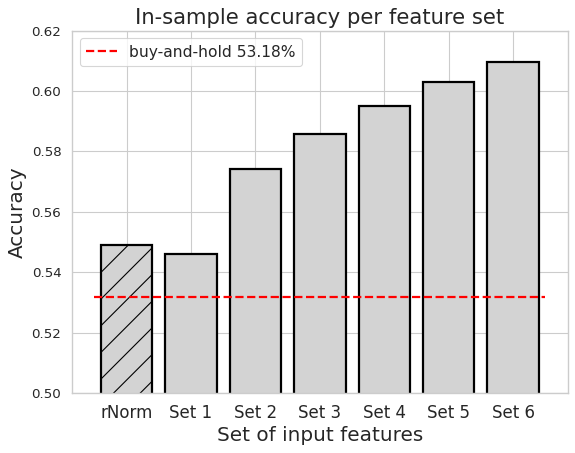}
    \caption{Average in-sample accuracy per feature set.}
    \label{fig:accuracyin}
  \end{minipage}
  \hfill
  \begin{minipage}[b]{0.47\textwidth}
    \includegraphics[width=6.8cm]{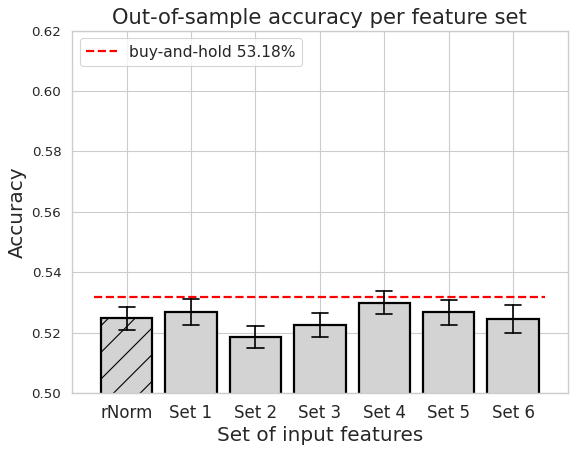}
    \caption{Average out-of-sample accuracy per feature set.}
    \label{fig:accuracyout}
  \end{minipage}
\end{figure}

\begin{figure}[ht]
  \centering
  \begin{minipage}[b]{0.47\textwidth}
    \includegraphics[width=6.8cm]{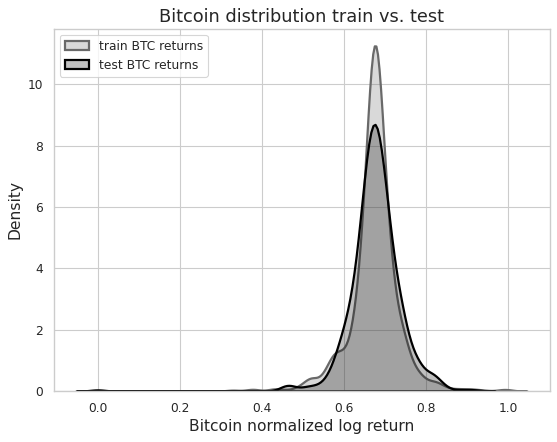}
    \caption{Distributions of returns in train and test set.}
    \label{fig:distribution}
  \end{minipage}
  \hfill
  \begin{minipage}[b]{0.47\textwidth}
    \includegraphics[width=6.8cm]{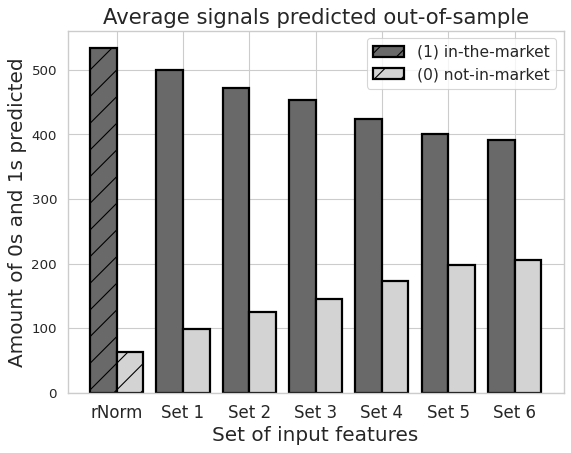}
    \caption{Average predicted signals per feature set.}
    \label{fig:signals}
  \end{minipage}
\end{figure}

The in-sample-accuracy in the training set illustrated in Fig. \ref{fig:accuracyin} improves monotonically with additional inputs from 54.6\% in feature set 1 up to 61\% in feature set 6. This increase is expected because more input features enable the network to make more dependencies. However, this can also lead to over-sampling if the out-of-sample accuracy is not increasing with additional features. We can not observe a monotonically out-of-sample accuracy increase in Fig. \ref{fig:accuracyout} in contrast to the the training set. As mentioned above, it appears that the additional input features lead to overfitting rather than providing real added value. In feature set 4, where EUR/USD is added, an increase in accuracy is recorded, however, accuracy drops again when the 10Y Treasury Yield is added. All in all, we cannot identify any clear added value at this stage. Furthermore, the accuracy of the rNorm feature set is similar to that of the other feature sets. The 95\% confidence intervals show that the true mean of the out-of-sample accuracy should be in these intervals with a probability of 95\%. The lower limit of the intervals is above 50\% for all feature sets. This might show that there is indeed a possible outperformance. However, it is important to note that in the test dataset (out-of-sample) there are 318 ones or positive signals (53.2\%) out of the total 598 observations that are in the sample. Consequently, a buy-and-hold strategy would have an accuracy of 53.2\%. As we can see in Fig. \ref{fig:signals}, the model predicts a positive signal more often than a negative one. Therefore, the accuracy of at least 51.9\% that is achieved by the neural network is probably not due to the model, but rather to the data distribution itself.
Barandela et al. \cite{imbalance_2004} describe the problem of an imbalanced training dataset in supervised pattern recognition. The problem occurs when one class is overrepresented in the training set while the other class is less frequent. Especially with time series that are usually rather noisy, this can lead to the model generally predicting the overrepresented class more often. If the test dataset then also has a similar imbalance, the prediction accuracy will likely be higher without the features contributing anything. In the training set used in this study class 1 (positive signal) is overrepresented with 55.96\%. Therefore it is likely that due to this small imbalance, the model will also predict class 1 more often than class 0. In the testing set, class 1 accounts for 53.18\%. Therefore, it can be assumed that the out-of-sample prediction accuracy will most likely be above 50\%, which applies to our case. Fig. \ref{fig:distribution} shows that the distributions of Bitcoin returns in the training and test datasets are not exactly the same. Fig. \ref{fig:signals} shows that the number of predicted positive signals decreases with each added feature.

\newpage

\begin{figure}[ht]
  \centering
  \begin{minipage}[b]{0.47\textwidth}
    \includegraphics[width=6.8cm]{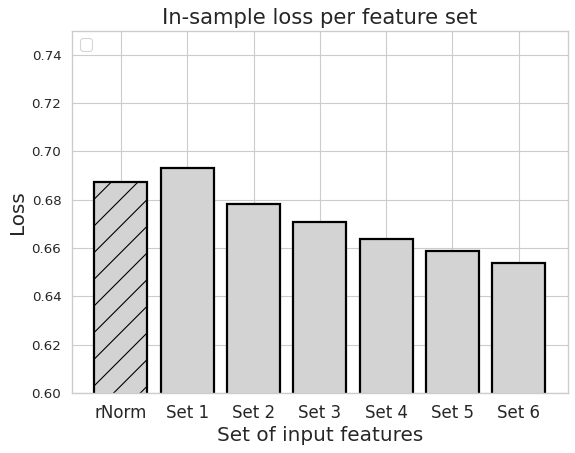}
    \caption{Average final in-sample loss.}
    \label{fig:lossin}
  \end{minipage}
  \hfill
  \begin{minipage}[b]{0.47\textwidth}
    \includegraphics[width=6.8cm]{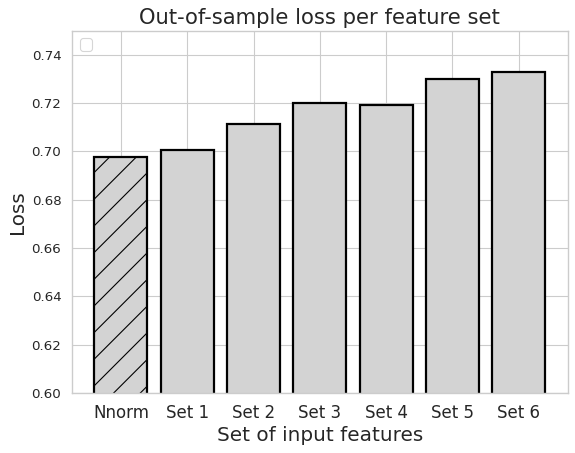}
    \caption{Average final out-of-sample loss.}
    \label{fig:lossout}
  \end{minipage}
\end{figure}

The loss is described by the binary cross-entropy function \ref{eqn:crossentropy}. In order to compare the individual feature sets, the last loss value of each iteration is stored and then the average of the 50 iterations is taken. The in-sample loss decreases with more input features as Fig. \ref{fig:lossin} shows. The reason is the same as with the in-sample accuracy, that more features give more scope to recognize dependencies. This makes the "punishment" value for errors smaller. On the other hand, in Fig. \ref{fig:lossout} the loss in the test dataset increases with almost every feature added, which is a sign of overfitting.

\subsection{Performance}

\begin{figure}[ht]
  \centering
  \begin{minipage}[b]{0.47\textwidth}
    \includegraphics[width=6.8cm]{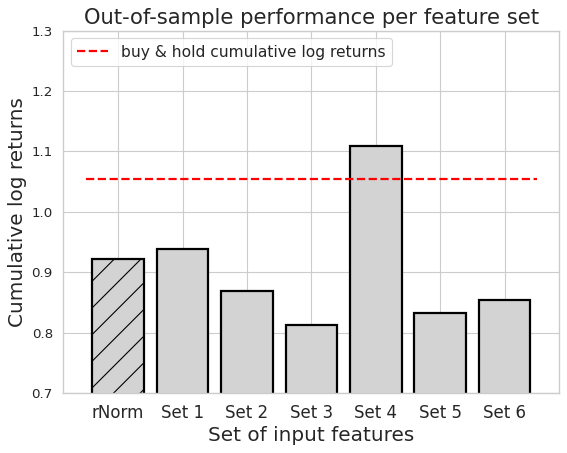}
    \caption{Average cumulative log returns.}
    \label{fig:cumsum}
  \end{minipage}
  \hfill
  \begin{minipage}[b]{0.47\textwidth}
    \includegraphics[width=6.8cm]{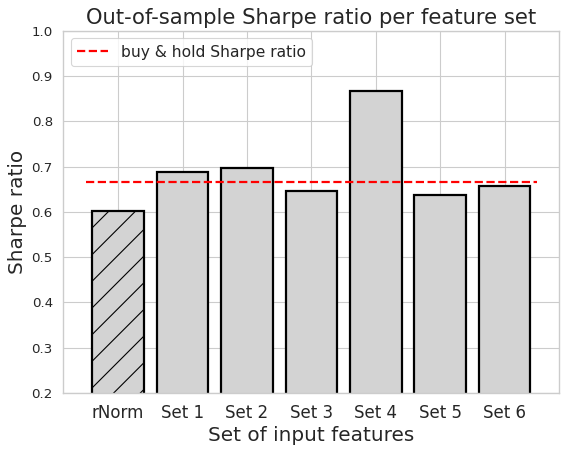}
    \caption{Average Sharpe ratios.}
    \label{fig:sharperatios}
  \end{minipage}
\end{figure}

The effective (buy-and-hold) cumulative sum of log returns of Bitcoin in the test dataset is 1.054. The cumulative sum of log returns achieved with the different feature sets is calculated with equation \ref{eqn:cumsum}. Each feature set generates a vector of signals (0s and 1s) in the length of the test dataset. For simplicity reasons, we use a long-only strategy for the experiment. It is hereby defined that 0 stands for “not in the market” or “not invested” and 1 stands for “in the market” or “invested”. Therefore when calculating the performance, only log returns that occurred at a positive signal will be cumulated and all log returns at a negative signal are ignored. For comparison, we take the average of all cumulative sums of log returns in each feature set (e.g. 50 iterations of feature set 1 then take the average). Fig. \ref{fig:cumsum} shows that only in feature set 4 is the achieved performance higher than it would be with a buy-and-hold approach. Furthermore, additional input features do not seem to linearly lead to better performance. As a second performance metric, the Sharpe ratio is calculated with equation \ref{eqn:sharperatio}. Fig. \ref{fig:sharperatios} shows, that the Sharpe ratio is sometimes better than with buy-and-hold, but this is probably also due to the fact that the strategy of the model is not always invested, and thus the volatility is lower. Both plots have a similar shape as the out-of-sample accuracy with the feature set 4 being significantly better than the others.\\

\begin{equation}
\label{eqn:cumsum}
    \text{Cumulative Log Returns} = \vec{P} \cdot \vec{R}
\end{equation}

\begin{equation}
\label{eqn:sharperatio}
    \text{Sharpe Ratio} = \frac{R_p}{\sigma_{p}} \cdot \sqrt{252}
\end{equation}

where:

\(\vec{P}\) is the vector of signals (0s \& 1s) predicted

\(\vec{R}\) is the vector of effective Bitcoin log returns during the test period

\(R_p\) is the average cumulative log returns over 50 iterations

\(\sigma_{p}\) is the average daily standard deviation over 50 iterations

\section{Discussion}

\textbf{General discussion}

As expected, the in-sample accuracy increases with each input feature added, as the model has more scope to explain the signals. However, the accuracy that is generated out-of-sample on the test dataset does not improve consistently. The feature set 4 has the highest accuracy of 53\% but in feature set 5 when the 10Y Treasury Yield was added, accuracy drops again, so the higher accuracy in feature set 4 appears to be random. This is also shown by the comparison with the benchmark rNorm, which achieves a higher out-of-sample predictive accuracy than the other feature sets in 3 out of 6 cases. Furthermore, a buy-and-hold strategy would be correct 53.2\% of the time during the test sample period. The fact that the out-of-sample accuracy is not improving the same way as the in-sample accuracy is a sign that additional inputs lead to overfitting and, likely, they do not have a dependency on the target. We can also see the same behavior with the loss on the training (decreasing) and test set (increasing). Additional input features seem to make the model more active. It predicts more negative Bitcoin signals with each input added. The random feature set rNorm predicts a positive signal 89\% of the time, while the largest feature set 6 predicts a positive signal only 66\% of the time. Exactly why this is the case would need to be investigated in a follow-up study.\\
The performance (cumulative log returns) is always worse than buy-and-hold, except again for the feature set 4, which is also better than the other feature sets in terms of accuracy. However, the performance of feature set 5 and 6 are again significantly worse than that of feature set 4. The Sharpe ratios are partly better or about the same as buy-and-hold, but with every feature set, the model is not always invested, which lowers the volatility compared to buy-and-hold. The fact that the feature set rNorm consisting of random values is not consistently outperformed also suggests that the features chosen in this study are not able to explain the Bitcoin price.\\

\textbf{Research questions}  

\textbf{RC 1:} Do additional input features improve the prediction accuracy?

No consistent improvement in prediction accuracy is recorded in the test samples.\\

\textbf{RC 2:} Is the Bitcoin market efficient as per the weak form of the EMH?

Since buy-and-hold is not outperformed with any of the feature sets in terms of accuracy, the Bitcoin market appears to be weakly efficient under the chosen scenario.\\

These two assertions only refer to the selected test period, the network architecture, the input features, the number of feature lags, and hyperparameters for this experiment. If these parameters are changed, other results will be the consequence. Also, other combinations of input features may lead to other results.

\section{Conclusion}

Since much of this experiment is based on chance, we encourage further studies on this topic. Models based on and also tested on past data are unfortunately never 100\% accurate or meaningful, as the author can adjust the constellation to produce results in the test sets that are favorable to him. We attempted to address this problem by specifying a model architecture and then averaging the values of the metrics over 50 iterations in Table \ref{table:6} before evaluating the individual feature sets. The results obtained from this study lead us to the partial conclusion that weak market inefficiencies for Bitcoin cannot be detected using neural networks and the selected asset classes as input.

\newpage

\addcontentsline{toc}{section}{References}
\bibliographystyle{IEEEtran_edited2}    
\bibliography{references}

\begin{thebibliography}{10}
\providecommand{\url}[1]{#1}
\csname url@samestyle\endcsname
\providecommand{\newblock}{\relax}
\providecommand{\bibinfo}[2]{#2}
\providecommand{\BIBentrySTDinterwordspacing}{\spaceskip=0pt\relax}
\providecommand{\BIBentryALTinterwordstretchfactor}{4}
\providecommand{\BIBentryALTinterwordspacing}{\spaceskip=\fontdimen2\font plus
\BIBentryALTinterwordstretchfactor\fontdimen3\font minus
  \fontdimen4\font\relax}
\providecommand{\BIBforeignlanguage}[2]{{%
\expandafter\ifx\csname l@#1\endcsname\relax
\typeout{** WARNING: IEEEtran.bst: No hyphenation pattern has been}%
\typeout{** loaded for the language `#1'. Using the pattern for}%
\typeout{** the default language instead.}%
\else
\language=\csname l@#1\endcsname
\fi
#2}}
\providecommand{\BIBdecl}{\relax}
\BIBdecl

\bibitem{malkiel_2005}
\BIBentryALTinterwordspacing
B.~G. Malkiel, ``Reflections on the efficient market hypothesis: 30 years
  later,'' \emph{Financial Review}, vol.~40, no.~1, pp. 1--9, Feb. 2005.
\BIBentrySTDinterwordspacing

\bibitem{Wildi_2019}
\BIBentryALTinterwordspacing
M.~Wildi and N.~A. Bundi, ``Bitcoin and market-(in)efficiency : a systematic
  time series approach,'' \emph{Digital Finance}, vol.~1, pp. 47--65, Nov.
  2019.
\BIBentrySTDinterwordspacing

\bibitem{Huang_2019}
\BIBentryALTinterwordspacing
J.-Z. Huang, W.~Huang, and J.~Ni, ``Predicting bitcoin returns using
  high-dimensional technical indicators,'' \emph{The Journal of Finance and
  Data Science}, vol.~5, pp. 140--155, Sep. 2019.
\BIBentrySTDinterwordspacing

\bibitem{Mudassir_2020}
\BIBentryALTinterwordspacing
M.~Mudassir, S.~Bennbaia, D.~Unal, and M.~Hammoudeh, ``Time-series forecasting
  of bitcoin prices using high-dimensional features: a machine learning
  approach,'' \emph{Neural Computing and Applications}, Jul. 2020.
\BIBentrySTDinterwordspacing

\bibitem{Jaquart_2021}
\BIBentryALTinterwordspacing
P.~Jaquart, D.~Dann, and C.~Weinhardt, ``Short-term bitcoin market prediction
  via machine learning,'' \emph{The Journal of Finance and Data Science},
  vol.~7, pp. 45--66, Nov. 2021.
\BIBentrySTDinterwordspacing

\bibitem{Chen_2020}
\BIBentryALTinterwordspacing
Z.~Chen, C.~Li, and W.~Sun, ``Bitcoin price prediction using machine learning:
  An approach to sample dimension engineering,'' \emph{Journal of Computational
  and Applied Mathematics}, vol. 365, Feb. 2020.
\BIBentrySTDinterwordspacing

\bibitem{Fama_1965a}
\BIBentryALTinterwordspacing
E.~F. Fama, ``The behavior of stock-market prices,'' \emph{The Journal of
  Business}, vol.~38, no.~1, pp. 34--105, Jan. 1965.
\BIBentrySTDinterwordspacing

\bibitem{Fama_1965b}
\BIBentryALTinterwordspacing
E.~F. Fama, ``Random walks in stock market prices,'' \emph{Financial Analysts
  Journal}, vol.~21, no.~5, pp. 55--59, Sep. 1965.
\BIBentrySTDinterwordspacing

\bibitem{Samuelson_1965}
\BIBentryALTinterwordspacing
P.~A. Samuelson, ``Rational theory of warrant pricing,'' \emph{Industrial
  Management Review}, vol.~6, no.~2, pp. 13--39, 1965.
\BIBentrySTDinterwordspacing

\bibitem{Fama_1970}
\BIBentryALTinterwordspacing
B.~G. Malkiel and E.~F. Fama, ``Efficient capital markets: A review of theory
  and empirical work,'' \emph{The Journal of Finance}, vol.~25, no.~2, pp.
  383--417, May 1970.
\BIBentrySTDinterwordspacing

\bibitem{EMH_limitations_2015}
\BIBentryALTinterwordspacing
E.~Sinclair, \emph{"The Efficient Market Hypothesis and Its Limitations," in
  Positional Option Trading}.\hskip 1em plus 0.5em minus 0.4em\relax John Wiley
  \& Sons, Ltd, Sep. 2015, ch.~2, pp. 11--27.
\BIBentrySTDinterwordspacing

\bibitem{munger1999}
\BIBentryALTinterwordspacing
C.~Munger. Markets are fairly efficient; berkshire hathaway 1999 annual
  meeting. Accessed: 2022-05-31. [Online]. Available:
  \url{https://buffett.cnbc.com/video/1999/05/03/afternoon-session---1999-berkshire-hathaway-annual-meeting.html}
\BIBentrySTDinterwordspacing

\bibitem{Stiglitz_1980}
\BIBentryALTinterwordspacing
S.~J. Grossman and J.~E. Stiglitz, ``On the impossibility of informationally
  efficient markets,'' \emph{The American Economic Review}, vol.~70, no.~3, pp.
  393--408, Jun. 1980.
\BIBentrySTDinterwordspacing

\bibitem{AMH_2004}
\BIBentryALTinterwordspacing
A.~W. Lo, ``The adaptive markets hypothesis: Market efficiency from an
  evolutionary perspective,'' \emph{Journal of Portfolio Management}, vol.~30,
  pp. 15--29, Oct. 2004.
\BIBentrySTDinterwordspacing

\bibitem{Efficiency_Rationality_2010}
\BIBentryALTinterwordspacing
J.~R. Boatright and N.-H. Hsieh, \emph{"Efficiency and Rationality," in Finance
  Ethics}.\hskip 1em plus 0.5em minus 0.4em\relax John Wiley \& Sons, Ltd,
  2010, ch.~4, pp. 63--83.
\BIBentrySTDinterwordspacing

\bibitem{nakamoto}
\BIBentryALTinterwordspacing
S.~Nakamoto, ``Bitcoin: A peer-to-peer electronic cash system,'' 2008.
  [Online]. Available: \url{https://bitcoin.org/bitcoin.pdf}
\BIBentrySTDinterwordspacing

\bibitem{garcia_2014}
\BIBentryALTinterwordspacing
D.~Garcia, C.~Tessone, P.~Mavrodiev, and N.~Perony, ``The digital traces of
  bubbles: Feedback cycles between socio-economic signals in the bitcoin
  economy,'' \emph{J. R. Soc. Interface}, vol.~11, Aug. 2014.
\BIBentrySTDinterwordspacing

\bibitem{Hencic_2015}
\BIBentryALTinterwordspacing
A.~Hencic and C.~Gourieroux, ``Noncausal autoregressive model in application to
  bitcoin/usd exchange rates,'' \emph{Studies in Computational Intelligence},
  vol. 583, pp. 17--40, Jan. 2014.
\BIBentrySTDinterwordspacing

\bibitem{Sapuric_2014}
\BIBentryALTinterwordspacing
S.~Sapuric and A.~Kokkinaki, ``\BIBforeignlanguage{English}{Bitcoin is
  volatile! isn{\textquoteright}t that right?}''
  \emph{\BIBforeignlanguage{English}{Lecture Notes in Business Information
  Processing}}, vol. 183, pp. 255--265, Oct. 2014.
\BIBentrySTDinterwordspacing

\bibitem{Kristoufek_2015}
\BIBentryALTinterwordspacing
L.~Kristoufek, ``What are the main drivers of the bitcoin price? evidence from
  wavelet coherence analysis,'' \emph{PLOS ONE}, vol.~10, no.~4, pp. 1--15,
  Apr. 2015.
\BIBentrySTDinterwordspacing

\bibitem{osterrieder_statistical_2017_lorenz}
\BIBentryALTinterwordspacing
J.~Osterrieder and J.~Lorenz, ``A statistical risk assessment of bitcoin and
  its extreme tail behavior,'' \emph{Annals of Financial Economics}, vol.~12,
  no.~01, p. 1750003, Apr. 2017.
\BIBentrySTDinterwordspacing

\bibitem{Osterrieder}
\BIBentryALTinterwordspacing
S.~Chan, J.~Chu, S.~Nadarajah, and J.~Osterrieder, ``A statistical analysis of
  cryptocurrencies,'' \emph{Journal of Risk and Financial Management}, vol.~10,
  no.~2, May 2017.
\BIBentrySTDinterwordspacing

\bibitem{Enoksen_2020}
\BIBentryALTinterwordspacing
F.~Enoksen, C.~Landsnes, K.~Lučivjanská, and P.~Molnár, ``Understanding risk
  of bubbles in cryptocurrencies,'' \emph{Journal of Economic Behavior \&
  Organization}, vol. 176, pp. 129--144, Aug. 2020.
\BIBentrySTDinterwordspacing

\bibitem{Urquhart_2016}
\BIBentryALTinterwordspacing
A.~Urquhart, ``The inefficiency of bitcoin,'' \emph{Economics Letters}, vol.
  148, pp. 80--82, Nov. 2016.
\BIBentrySTDinterwordspacing

\bibitem{Nadarajah_2017}
\BIBentryALTinterwordspacing
S.~Nadarajah and J.~Chu, ``On the inefficiency of bitcoin,'' \emph{Economics
  Letters}, vol. 150, pp. 6--9, Jan. 2017.
\BIBentrySTDinterwordspacing

\bibitem{Bariviera_2017}
\BIBentryALTinterwordspacing
A.~F. Bariviera, ``The inefficiency of bitcoin revisited: A dynamic approach,''
  \emph{Economics Letters}, vol. 161, pp. 1--4, Dec. 2017.
\BIBentrySTDinterwordspacing

\bibitem{Brauneis_2018}
\BIBentryALTinterwordspacing
A.~Brauneis and R.~Mestel, ``Price discovery of cryptocurrencies: Bitcoin and
  beyond,'' \emph{Economics Letters}, vol. 165, pp. 58--61, Apr. 2018.
\BIBentrySTDinterwordspacing

\bibitem{Tiwari_2018}
\BIBentryALTinterwordspacing
A.~K. Tiwari, R.~Jana, D.~Das, and D.~Roubaud, ``Informational efficiency of
  bitcoin—an extension,'' \emph{Economics Letters}, vol. 163, pp. 106--109,
  Feb. 2018.
\BIBentrySTDinterwordspacing

\bibitem{ALYAHYAEE_2018}
\BIBentryALTinterwordspacing
K.~H. Al-Yahyaee, W.~Mensi, and S.-M. Yoon, ``Efficiency, multifractality, and
  the long-memory property of the bitcoin market: A comparative analysis with
  stock, currency, and gold markets,'' \emph{Finance Research Letters},
  vol.~27, pp. 228--234, Dec. 2018.
\BIBentrySTDinterwordspacing

\bibitem{Sensoy_2018}
\BIBentryALTinterwordspacing
A.~Sensoy, ``The inefficiency of bitcoin revisited: A high-frequency analysis
  with alternative currencies,'' \emph{Finance Research Letters}, vol.~28, pp.
  68--73, Mar. 2019.
\BIBentrySTDinterwordspacing

\bibitem{vidal_2018}
\BIBentryALTinterwordspacing
D.~Vidal-Tomás and A.~Ibañez, ``Semi-strong efficiency of bitcoin,''
  \emph{Finance Research Letters}, vol.~27, pp. 259--265, Dec. 2018.
\BIBentrySTDinterwordspacing

\bibitem{Khuntia_2018}
\BIBentryALTinterwordspacing
S.~Khuntia and J.~Pattanayak, ``Adaptive market hypothesis and evolving
  predictability of bitcoin,'' \emph{Economics Letters}, vol. 167, pp. 26--28,
  Jun. 2018.
\BIBentrySTDinterwordspacing

\bibitem{Shihao_2020}
\BIBentryALTinterwordspacing
S.~Gu, B.~Kelly, and D.~Xiu, ``Empirical asset pricing via machine learning,''
  \emph{The Review of Financial Studies}, vol.~33, p. 2223–2273, Mai. 2020.
\BIBentrySTDinterwordspacing

\bibitem{Guanhao_2020}
\BIBentryALTinterwordspacing
G.~Feng and S.~Giglio, ``Taming the factor zoo: A test of new factors,''
  \emph{The Journal of Finance}, vol.~75, pp. 1327--1370, Jun. 2020.
\BIBentrySTDinterwordspacing

\bibitem{Belloni}
\BIBentryALTinterwordspacing
A.~Belloni, V.~Chernozhukov, and C.~Hansen, ``Inference on treatment effects
  after selection among high-dimensional controls,'' \emph{The Review of
  Economic Studies}, vol.~81, no. 2 (287), pp. 608--650, 2014.
\BIBentrySTDinterwordspacing

\bibitem{coinbase_closing}
\BIBentryALTinterwordspacing
CoinMarketCap. Market open and close times. Accessed: 2022-05-24. [Online].
  Available:
  \url{https://support.coinmarketcap.com/hc/en-us/articles/360016193231-Market-Open-and-Close-Times#:~:text=The\%20market\%20opens\%20at\%2012,UTC\%20time\%20unless\%20otherwise\%20specified}
\BIBentrySTDinterwordspacing

\bibitem{NYSE_closing}
\BIBentryALTinterwordspacing
NYSE. Holidays \& trading hours. Accessed: 2022-05-24. [Online]. Available:
  \url{https://www.nyse.com/markets/hours-calendars}
\BIBentrySTDinterwordspacing

\bibitem{EURUSD_YF}
\BIBentryALTinterwordspacing
YahooFinance. Eur/usd (eurusd=x). Accessed: 2022-05-24. [Online]. Available:
  \url{https://finance.yahoo.com/quote/EURUSD=X/}
\BIBentrySTDinterwordspacing

\bibitem{Treasury_YF}
\BIBentryALTinterwordspacing
YahooFinance. Treasury yield 10 years (\^{}tnx). Accessed: 2022-05-24.
  [Online]. Available: \url{https://finance.yahoo.com/quote/%5ETNX/}
\BIBentrySTDinterwordspacing

\bibitem{xau_bloomberg}
Bloomberg. Philadelphia gold and silver index description. Accessed:
  2022-04-28.

\bibitem{Coinbase_Marketshare}
\BIBentryALTinterwordspacing
CoinMarketCap. Today's cryptocurrency prices by market cap. Accessed:
  2022-05-24. [Online]. Available: \url{https://coinmarketcap.com/de/}
\BIBentrySTDinterwordspacing

\bibitem{Coinbase_prices}
\BIBentryALTinterwordspacing
CoinMarketCap. How are prices calculated on coinmarketcap? Accessed:
  2022-05-24. [Online]. Available:
  \url{https://support.coinmarketcap.com/hc/en-us/articles/360015968632-How-are-prices-calculated-on-CoinMarketCap-}
\BIBentrySTDinterwordspacing

\bibitem{Kim_2020}
\BIBentryALTinterwordspacing
J.-M. Kim, S.-T. Kim, and S.~Kim, ``On the relationship of cryptocurrency price
  with us stock and gold price using copula models,'' \emph{Mathematics},
  vol.~8, no.~11, Sep. 2020.
\BIBentrySTDinterwordspacing

\bibitem{FAMA_Riskfactors_1993}
\BIBentryALTinterwordspacing
E.~F. Fama and K.~R. French, ``Common risk factors in the returns on stocks and
  bonds,'' \emph{Journal of Financial Economics}, vol.~33, pp. 3--56, Feb.
  1993.
\BIBentrySTDinterwordspacing

\bibitem{FTSE_Russell}
\BIBentryALTinterwordspacing
FTSE. Index factsheet: Russell 2000 index. Accessed: 2022-05-24. [Online].
  Available:
  \url{https://research.ftserussell.com/Analytics/FactSheets/temp/d5c141a6-bb80-4471-83f4-2bf18e473adb.pdf}
\BIBentrySTDinterwordspacing

\bibitem{Gold_CME}
\BIBentryALTinterwordspacing
CMEGroup. Gold futures and options. Accessed: 2022-05-31. [Online]. Available:
  \url{https://www.cmegroup.com/markets/metals/precious/gold.html}
\BIBentrySTDinterwordspacing

\bibitem{XAU_methodology}
\BIBentryALTinterwordspacing
NASDAQ. Phlx gold/silver index methodology. Accessed: 2022-05-24. [Online].
  Available: \url{https://indexes.nasdaqomx.com/docs/methodology_XAU.pdf}
\BIBentrySTDinterwordspacing

\bibitem{Nasdaq_methodology}
\BIBentryALTinterwordspacing
NASDAQ. Nasdaq index methodology guide. Accessed: 2022-05-24. [Online].
  Available:
  \url{https://indexes.nasdaqomx.com/docs/Nasdaq_Index_Methodology_Guide%20(1).pdf}
\BIBentrySTDinterwordspacing

\bibitem{Japkowicz_2001}
\BIBentryALTinterwordspacing
N.~Japkowicz, ``Supervised versus unsupervised binary-learning by feedforward
  neural networks,'' \emph{Machine Learning}, vol.~42, pp. 97--122, Jan. 2001.
\BIBentrySTDinterwordspacing

\bibitem{LSTM_1997}
\BIBentryALTinterwordspacing
S.~Hochreiter and J.~Schmidhuber, ``{Long Short-Term Memory},'' \emph{Neural
  Computation}, vol.~9, no.~8, pp. 1735--1780, Nov. 1997.
\BIBentrySTDinterwordspacing

\bibitem{MinMaxScaler}
\BIBentryALTinterwordspacing
Scikit-learn. Sklearn minmaxscaler. Accessed: 2022-05-24. [Online]. Available:
  \url{https://scikit-learn.org/stable/modules/generated/sklearn.preprocessing.MinMaxScaler.html}
\BIBentrySTDinterwordspacing

\bibitem{Github_kraehenbuehl}
\BIBentryALTinterwordspacing
M.~Kraehenbuehl. Code bachelorthesis neural network with bitcoin. Accessed:
  2022-05-29. [Online]. Available:
  \url{https://github.com/mikekraehenbuehl/BA_2022_Neural_Network_Bitcoin}
\BIBentrySTDinterwordspacing

\bibitem{keras}
\BIBentryALTinterwordspacing
Keras. Keras api reference. Accessed: 2022-05-24. [Online]. Available:
  \url{https://keras.io/api/}
\BIBentrySTDinterwordspacing

\bibitem{NN_architecture}
\BIBentryALTinterwordspacing
NN-SVG. Nn-architecture schematics. Accessed: 2022-05-24. [Online]. Available:
  \url{https://alexlenail.me/NN-SVG/}
\BIBentrySTDinterwordspacing

\bibitem{relu}
\BIBentryALTinterwordspacing
Kaggle. Rectified linear units {ReLU} in deep learning. Accessed: 2022-05-24.
  [Online]. Available:
  \url{https://www.kaggle.com/code/dansbecker/rectified-linear-units-relu-in-deep-learning/notebook}
\BIBentrySTDinterwordspacing

\bibitem{imbalance_2004}
\BIBentryALTinterwordspacing
R.~Barandela, R.~M. Valdovinos, J.~S. S{\'a}nchez, and F.~J. Ferri, ``The
  imbalanced training sample problem: Under or over sampling?'' in
  \emph{Structural, Syntactic, and Statistical Pattern Recognition}.\hskip 1em
  plus 0.5em minus 0.4em\relax Springer Verlag, Aug. 2004, pp. 806--814.
\BIBentrySTDinterwordspacing

\bibitem{maravall_1983}
\BIBentryALTinterwordspacing
A.~Maravall, ``An application of nonlinear time series forecasting,''
  \emph{Journal of Business \& Economic Statistics}, vol.~1, pp. 66--74, Jan.
  1983.
\BIBentrySTDinterwordspacing

\end{thebibliography}

\section{Appendix}

\subsection{Plots}
\textbf{Bitcoin autocorrelation}  \label{sec:autocorrelation}

\begin{figure}[h]
    \centering
    \includegraphics[width=10cm]{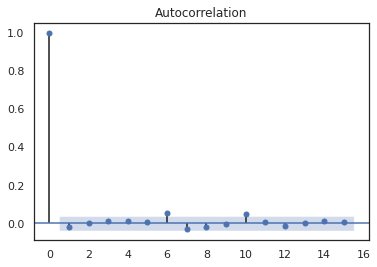}
    \caption{Plot shows autocorrelation for BTC at lag 6 \& 10.}
    \label{fig:autocorrelation}
\end{figure}

In Fig. \ref{fig:autocorrelation} only lag 6 and 10 appear to have a significant autocorrelation since the values are outside of the 95\% confidence interval. In the Bitcoin paper of Wildi and Bundi \cite{Wildi_2019} they also notice the autocorrelation at lag 6 and therefore use the first 6 lags as inputs for the neural network. However, despite this statistical significance, why should the Bitcoin price from 6 days or even 10 days ago have a dependency on today's price, whereas the prices of the last two days are completely irrelevant. Autocorrelation shows the amount of linear dependency in a time series. However, a lack of autocorrelation does not imply independence since it can not identify a possible nonlinearity \cite{maravall_1983}. Hence, a neural network with its nonlinear activation functions is a suitable predictive model for this study and we only select lag 1 as feature because intuitively it should be most relevant.

\newpage

\begin{figure}[ht]
    \centering
    \includegraphics[width=9cm]{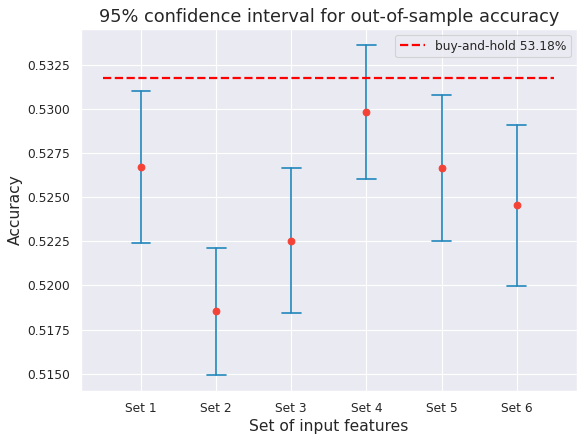}
    \caption{Confidence intervals for the true mean of the accuracies.}
    \label{fig:fig1}
\end{figure}

\newpage
\subsection{Results of individual feature sets} \label{sec:results}

\subsubsection{ETH same day (to test model performance)} \label{sec:ETH}

\begin{figure}[h]
  \centering
  \begin{minipage}[b]{0.47\textwidth}
    \includegraphics[width=6.8cm]{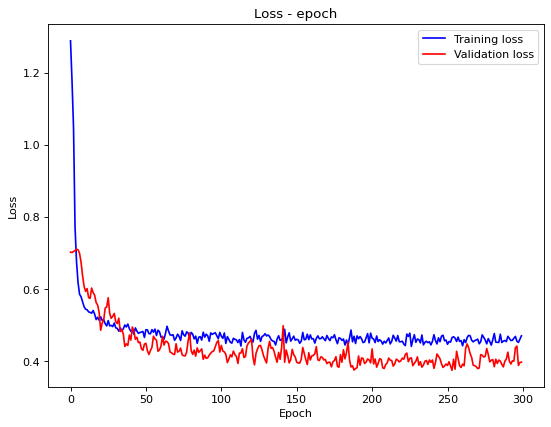}
    \caption{Train/Val. loss history of first iteration.}
  \end{minipage}
  \hfill
  \begin{minipage}[b]{0.47\textwidth}
    \includegraphics[width=6.8cm]{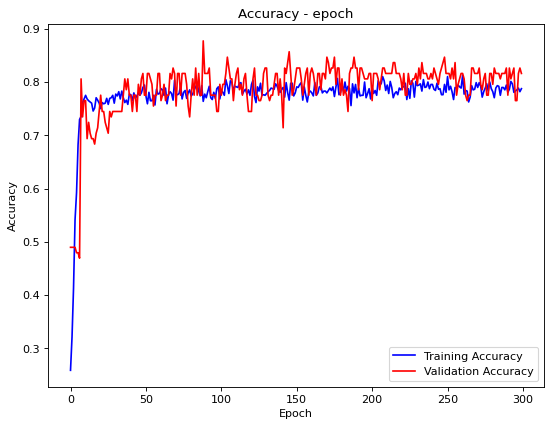}
    \caption{Train/Val. accuracy history of first iteration.}
  \end{minipage}
\end{figure}

\begin{figure}[h]
  \centering
  \begin{minipage}[b]{0.47\textwidth}
    \includegraphics[width=6.8cm]{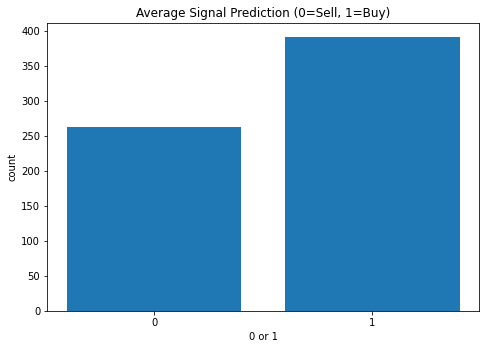}
    \caption{Average out-of-sample signals of 10 iterations.}
  \end{minipage}
  \hfill
  \begin{minipage}[b]{0.47\textwidth}
    \includegraphics[width=6.8cm]{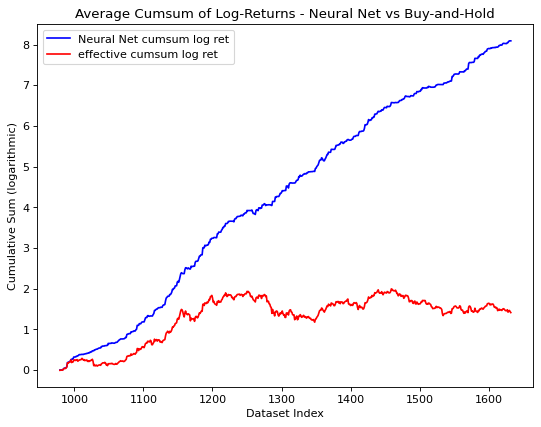}
    \caption{Average out-of-sample cumulative log returns of 10 iterations.}
  \end{minipage}
\end{figure}

\begin{figure}[h]
  \centering
  \begin{minipage}[b]{0.47\textwidth}
    \includegraphics[width=6.8cm]{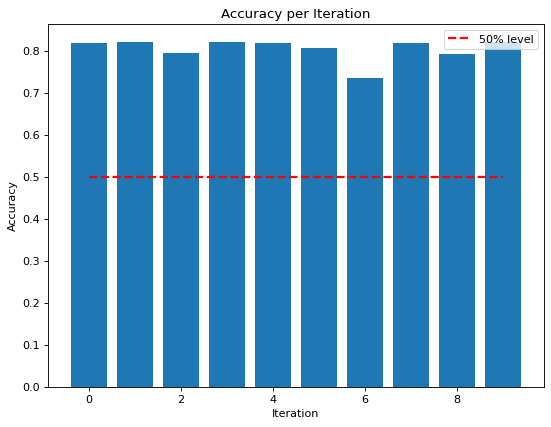}
    \caption{Out-of-sample accuracy per iteration.}
  \end{minipage}
  \hfill
  \begin{minipage}[b]{0.47\textwidth}
    \includegraphics[width=6.8cm]{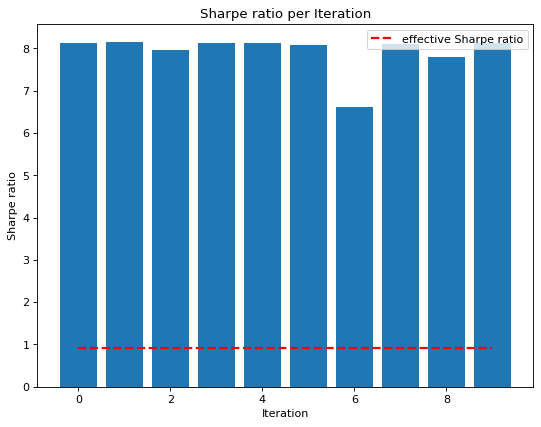}
    \caption{Out-of-sample Sharpe ratio per iteration.}
  \end{minipage}
\end{figure}

\newpage
\subsubsection{Feature set 0: random sample "rNorm"}

\begin{figure}[h]
  \centering
  \begin{minipage}[b]{0.47\textwidth}
    \includegraphics[width=7cm]{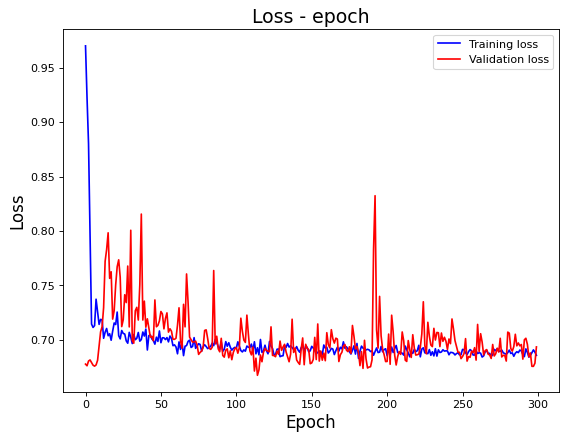}
    \caption{Train/Val. loss history of first iteration.}
  \end{minipage}
  \hfill
  \begin{minipage}[b]{0.47\textwidth}
    \includegraphics[width=7cm]{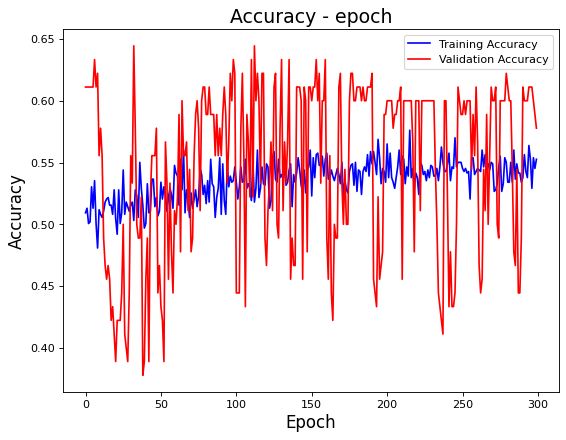}
    \caption{Train/Val. accuracy history of first iteration.}
  \end{minipage}
\end{figure}

\begin{figure}[h]
  \centering
  \begin{minipage}[b]{0.47\textwidth}
    \includegraphics[width=7cm]{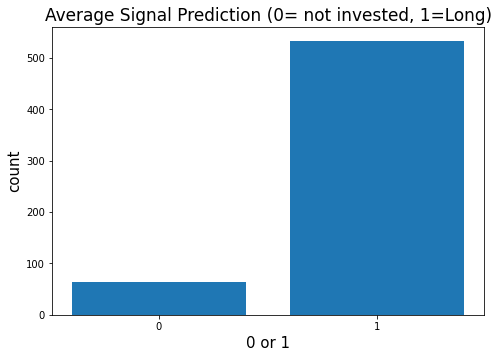}
    \caption{Average out-of-sample signals of 50 iterations.}
  \end{minipage}
  \hfill
  \begin{minipage}[b]{0.47\textwidth}
    \includegraphics[width=7cm]{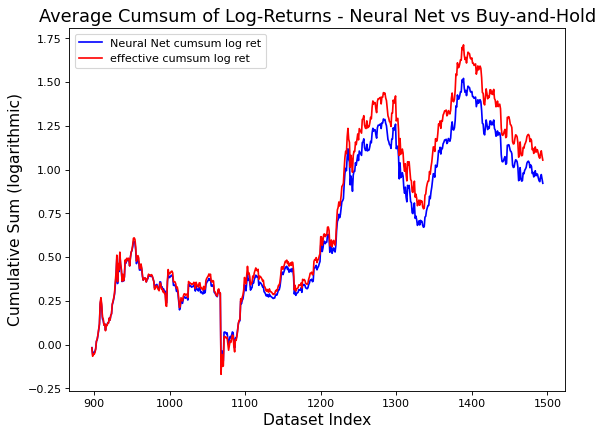}
    \caption{Average out-of-sample cumulative log returns of 50 iterations.}
  \end{minipage}
\end{figure}

\begin{figure}[h]
  \centering
  \begin{minipage}[b]{0.47\textwidth}
    \includegraphics[width=7cm]{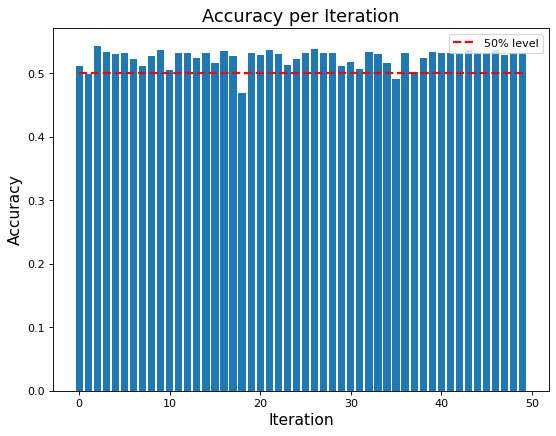}
    \caption{Out-of-sample accuracy per iteration.}
  \end{minipage}
  \hfill
  \begin{minipage}[b]{0.47\textwidth}
    \includegraphics[width=7cm]{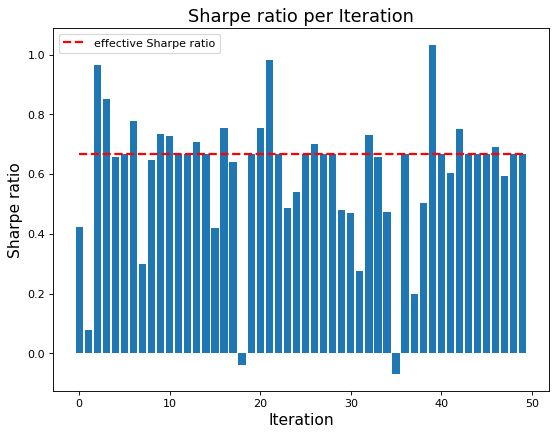}
    \caption{Out-of-sample Sharpe ratio per iteration.}
  \end{minipage}
\end{figure}

\newpage
\subsubsection{Feature set 1: Bitcoin}

\begin{figure}[h]
  \centering
  \begin{minipage}[b]{0.47\textwidth}
    \includegraphics[width=7cm]{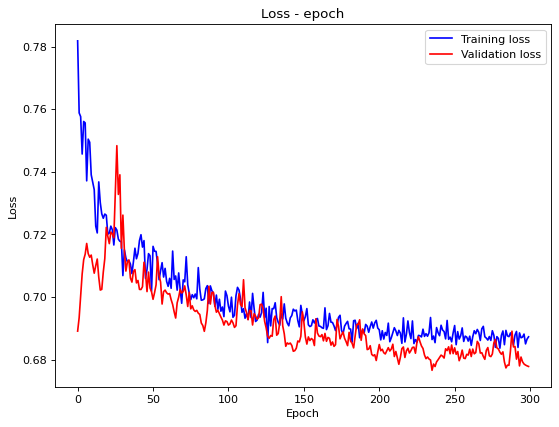}
    \caption{Train/Val. loss history of first iteration.}
  \end{minipage}
  \hfill
  \begin{minipage}[b]{0.47\textwidth}
    \includegraphics[width=7cm]{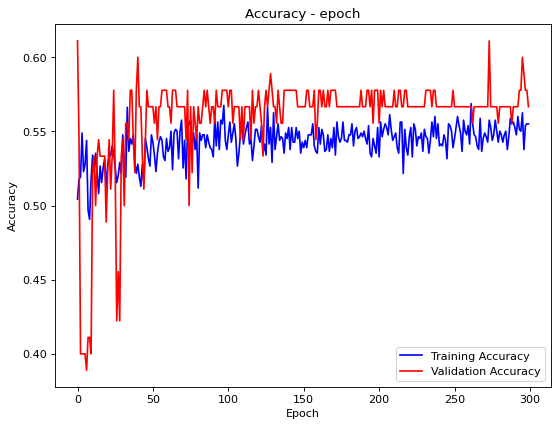}
    \caption{Train/Val. accuracy history of first iteration.}
  \end{minipage}
\end{figure}

\begin{figure}[h]
  \centering
  \begin{minipage}[b]{0.47\textwidth}
    \includegraphics[width=7cm]{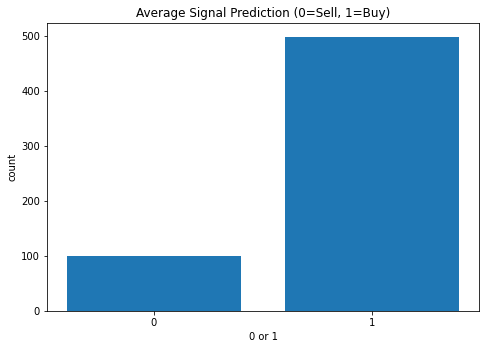}
    \caption{Average out-of-sample signals of 50 iterations.}
  \end{minipage}
  \hfill
  \begin{minipage}[b]{0.47\textwidth}
    \includegraphics[width=7cm]{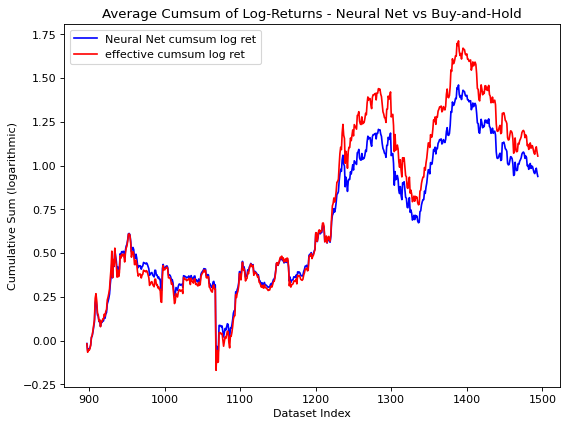}
    \caption{Average out-of-sample cumulative log returns of 50 iterations.}
  \end{minipage}
\end{figure}

\begin{figure}[h]
  \centering
  \begin{minipage}[b]{0.47\textwidth}
    \includegraphics[width=7cm]{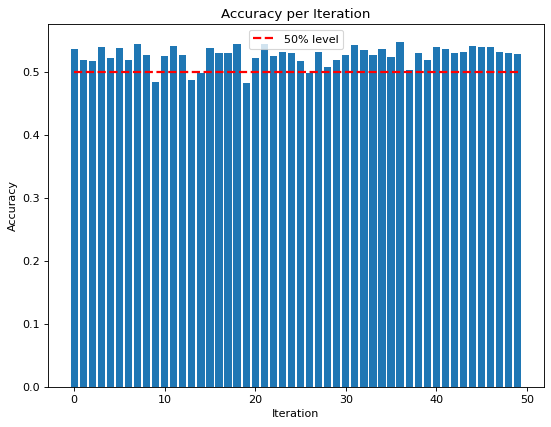}
    \caption{Out-of-sample accuracy per iteration.}
  \end{minipage}
  \hfill
  \begin{minipage}[b]{0.47\textwidth}
    \includegraphics[width=7cm]{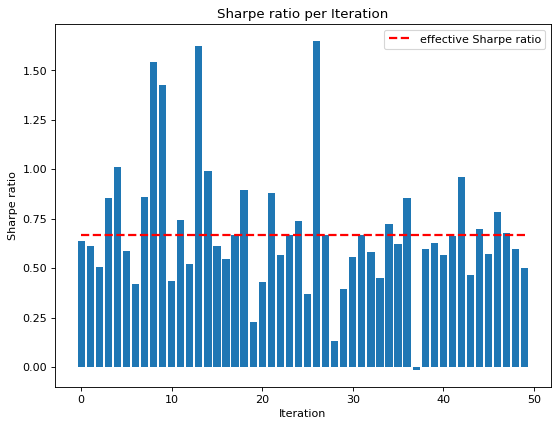}
    \caption{Out-of-sample Sharpe ratio per iteration.}
  \end{minipage}
\end{figure}

\newpage
\subsubsection{Feature set 2: Bitcoin + S\&P500}

\begin{figure}[h]
  \centering
  \begin{minipage}[b]{0.47\textwidth}
    \includegraphics[width=7cm]{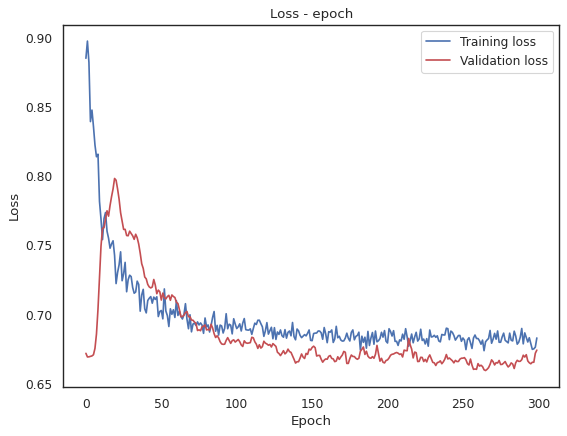}
    \caption{Train/Val. loss history of first iteration.}
  \end{minipage}
  \hfill
  \begin{minipage}[b]{0.47\textwidth}
    \includegraphics[width=7cm]{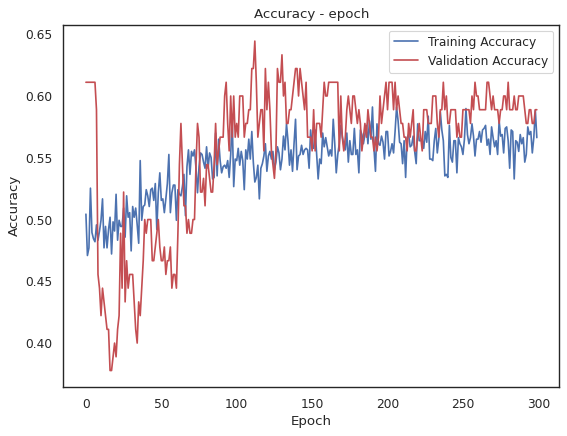}
    \caption{Train/Val. accuracy history of first iteration.}
  \end{minipage}
\end{figure}

\begin{figure}[h]
  \centering
  \begin{minipage}[b]{0.47\textwidth}
    \includegraphics[width=7cm]{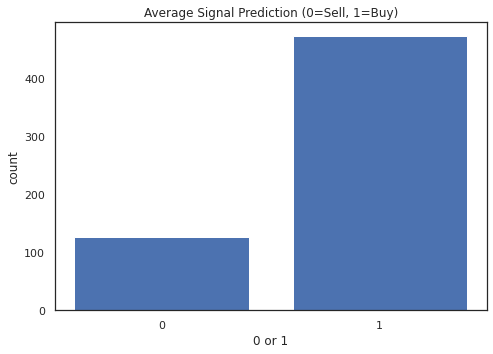}
    \caption{Average out-of-sample signals of 50 iterations.}
  \end{minipage}
  \hfill
  \begin{minipage}[b]{0.47\textwidth}
    \includegraphics[width=7cm]{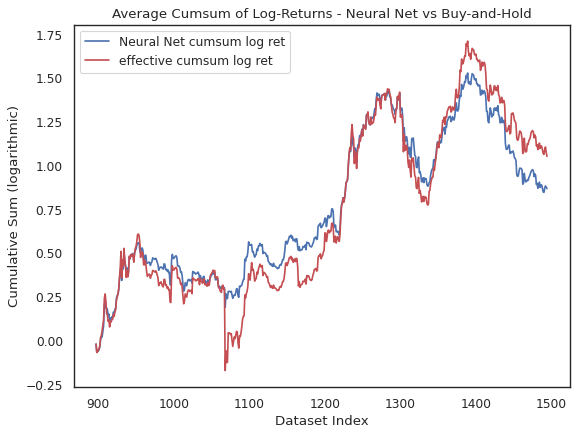}
    \caption{Average out-of-sample cumulative log returns of 50 iterations.}
  \end{minipage}
\end{figure}

\begin{figure}[h]
  \centering
  \begin{minipage}[b]{0.47\textwidth}
    \includegraphics[width=7cm]{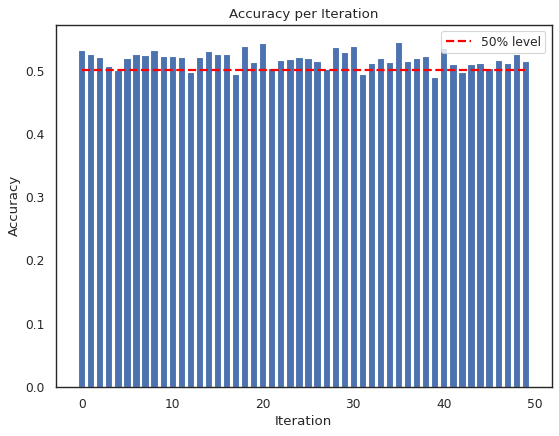}
    \caption{Out-of-sample accuracy per iteration.}
  \end{minipage}
  \hfill
  \begin{minipage}[b]{0.47\textwidth}
    \includegraphics[width=7cm]{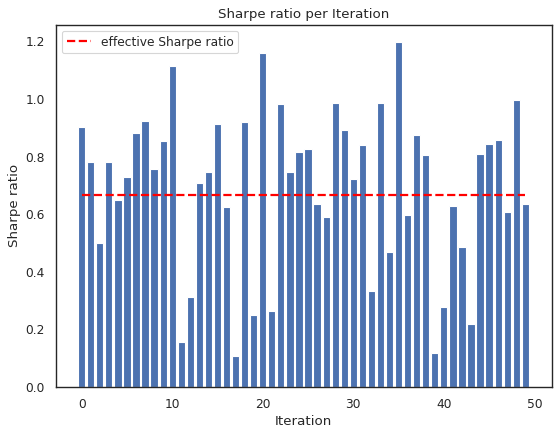}
    \caption{Out-of-sample Sharpe ratio per iteration.}
  \end{minipage}
\end{figure}

\newpage
\subsubsection{Feature set 3: Bitcoin + S\&P500 + Russell 2000}

\begin{figure}[h]
  \centering
  \begin{minipage}[b]{0.47\textwidth}
    \includegraphics[width=7cm]{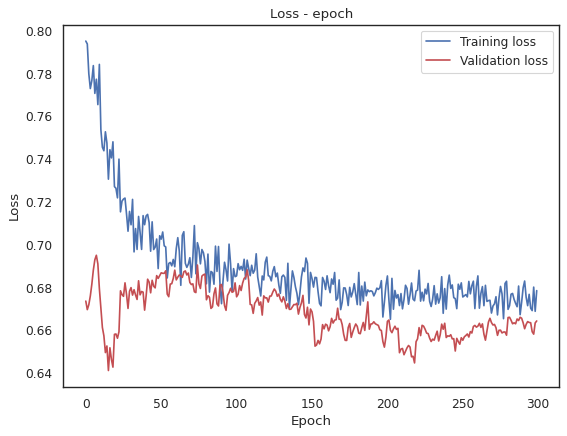}
    \caption{Train/Val. loss history of first iteration.}
  \end{minipage}
  \hfill
  \begin{minipage}[b]{0.47\textwidth}
    \includegraphics[width=7cm]{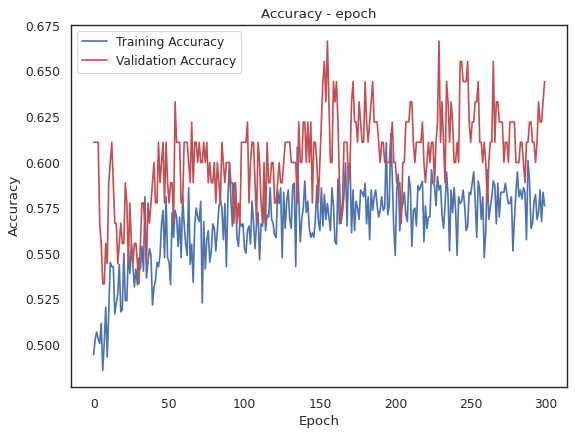}
    \caption{Train/Val. accuracy history of first iteration.}
  \end{minipage}
\end{figure}

\begin{figure}[h]
  \centering
  \begin{minipage}[b]{0.47\textwidth}
    \includegraphics[width=7cm]{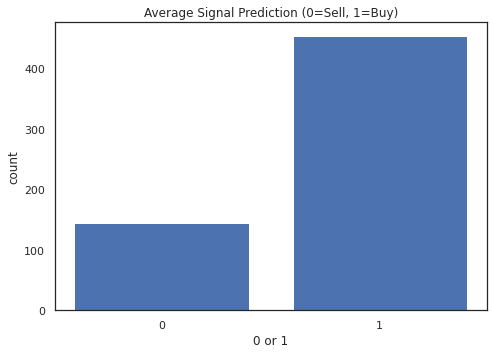}
    \caption{Average out-of-sample signals of 50 iterations.}
  \end{minipage}
  \hfill
  \begin{minipage}[b]{0.47\textwidth}
    \includegraphics[width=7cm]{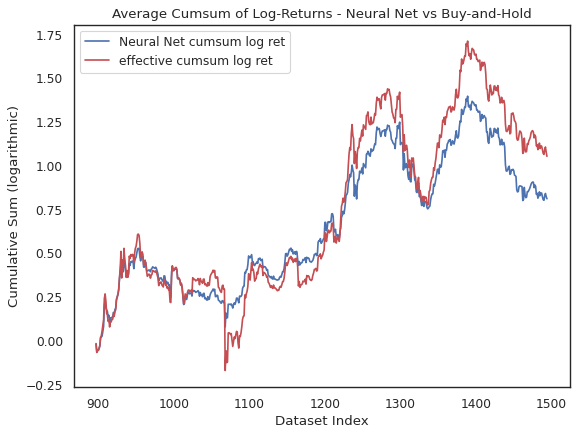}
    \caption{Average out-of-sample cumulative log returns of 50 iterations.}
  \end{minipage}
\end{figure}

\begin{figure}[h]
  \centering
  \begin{minipage}[b]{0.47\textwidth}
    \includegraphics[width=7cm]{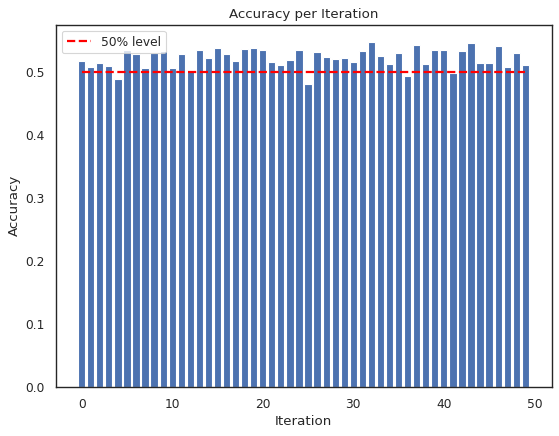}
    \caption{Out-of-sample accuracy per iteration.}
  \end{minipage}
  \hfill
  \begin{minipage}[b]{0.47\textwidth}
    \includegraphics[width=7cm]{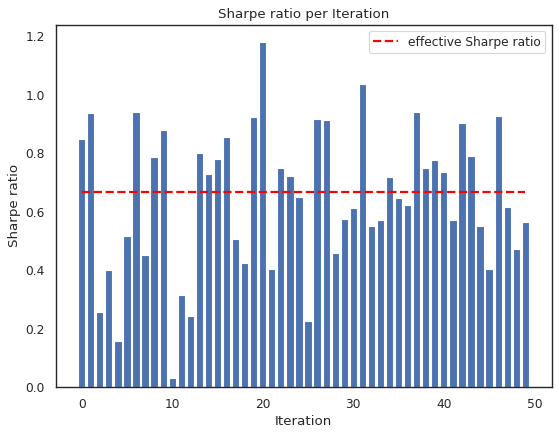}
    \caption{Out-of-sample Sharpe ratio per iteration.}
  \end{minipage}
\end{figure}

\newpage
\subsubsection{Feature set 4: Bitcoin + S\&P500 + Russell 2000 + EUR/USD}

\begin{figure}[h]
  \centering
  \begin{minipage}[b]{0.47\textwidth}
    \includegraphics[width=7cm]{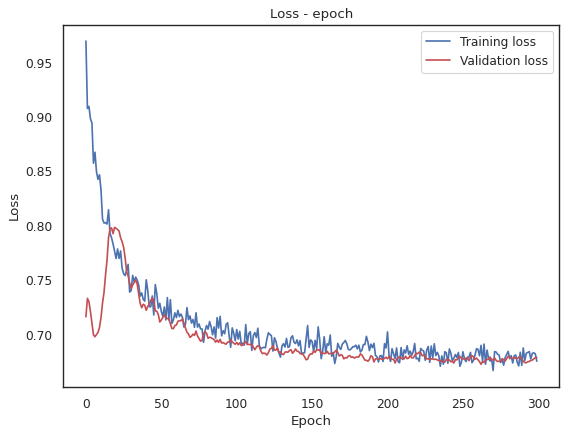}
    \caption{Train/Val. loss history of first iteration.}
  \end{minipage}
  \hfill
  \begin{minipage}[b]{0.47\textwidth}
    \includegraphics[width=7cm]{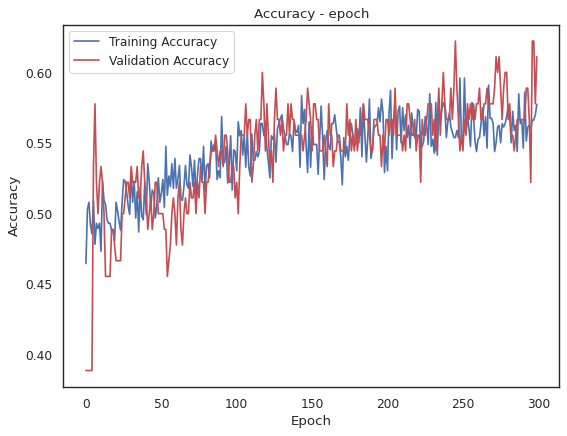}
    \caption{Train/Val. accuracy history of first iteration.}
  \end{minipage}
\end{figure}

\begin{figure}[h]
  \centering
  \begin{minipage}[b]{0.47\textwidth}
    \includegraphics[width=7cm]{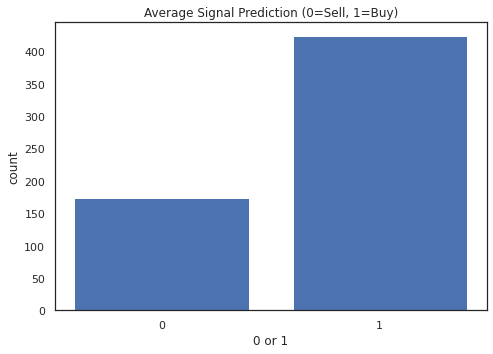}
    \caption{Average out-of-sample signals of 50 iterations.}
  \end{minipage}
  \hfill
  \begin{minipage}[b]{0.47\textwidth}
    \includegraphics[width=7cm]{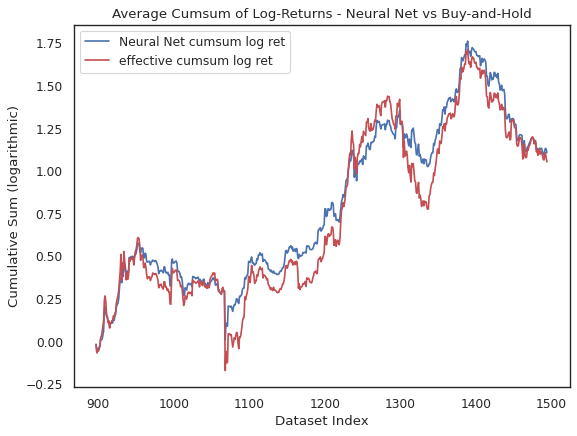}
    \caption{Average out-of-sample cumulative log returns of 50 iterations.}
  \end{minipage}
\end{figure}

\begin{figure}[h]
  \centering
  \begin{minipage}[b]{0.47\textwidth}
    \includegraphics[width=7cm]{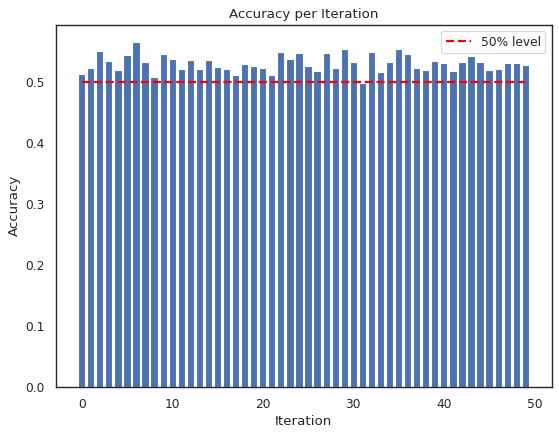}
    \caption{Out-of-sample accuracy per iteration.}
  \end{minipage}
  \hfill
  \begin{minipage}[b]{0.47\textwidth}
    \includegraphics[width=7cm]{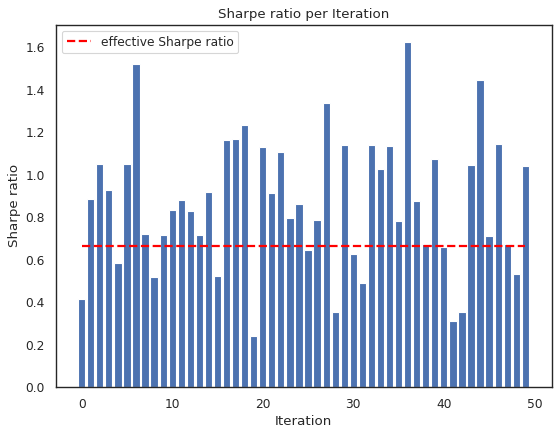}
    \caption{Out-of-sample Sharpe ratio per iteration.}
  \end{minipage}
\end{figure}

\newpage
\subsubsection{Feature set 5: Bitcoin + S\&P500 + Russell 2000 + EUR/USD + 10Y Treasury Yield}

\begin{figure}[h]
  \centering
  \begin{minipage}[b]{0.47\textwidth}
    \includegraphics[width=7cm]{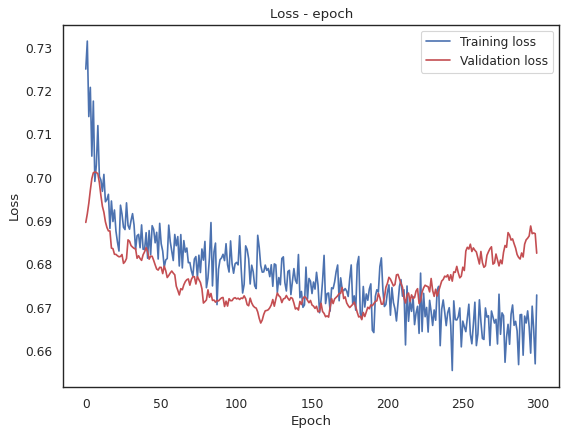}
    \caption{Train/Val. loss history of first iteration.}
  \end{minipage}
  \hfill
  \begin{minipage}[b]{0.47\textwidth}
    \includegraphics[width=7cm]{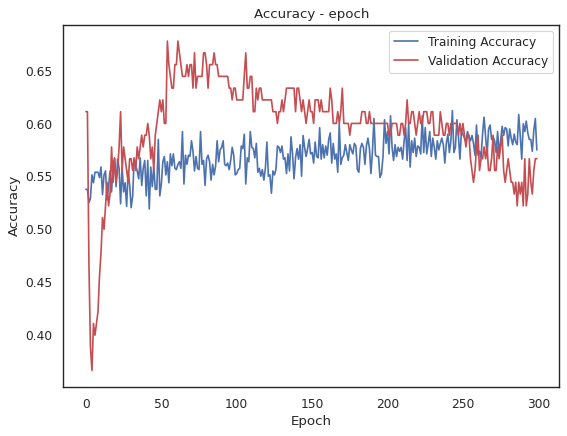}
    \caption{Train/Val. accuracy history of first iteration.}
  \end{minipage}
\end{figure}

\begin{figure}[h]
  \centering
  \begin{minipage}[b]{0.47\textwidth}
    \includegraphics[width=7cm]{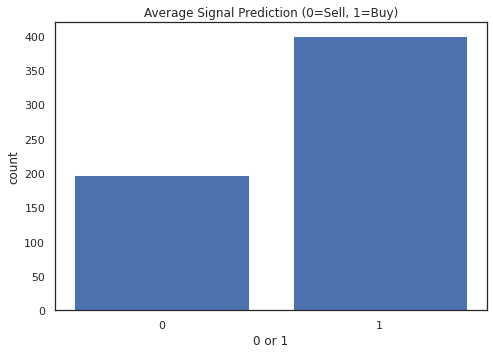}
    \caption{Average out-of-sample signals of 50 iterations.}
  \end{minipage}
  \hfill
  \begin{minipage}[b]{0.47\textwidth}
    \includegraphics[width=7cm]{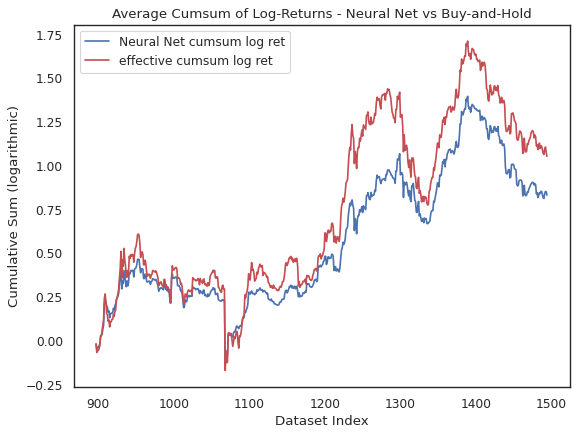}
    \caption{Average out-of-sample cumulative log returns of 50 iterations.}
  \end{minipage}
\end{figure}

\begin{figure}[h]
  \centering
  \begin{minipage}[b]{0.47\textwidth}
    \includegraphics[width=7cm]{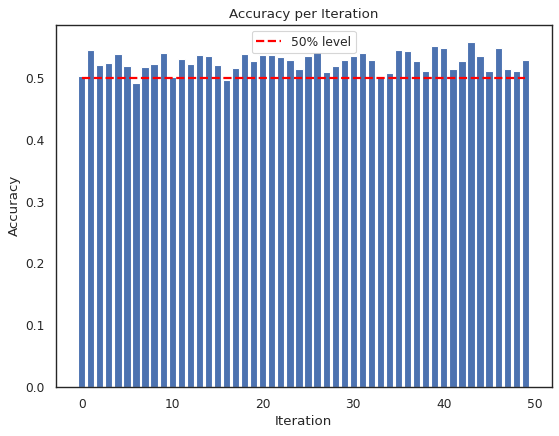}
    \caption{Out-of-sample accuracy per iteration.}
  \end{minipage}
  \hfill
  \begin{minipage}[b]{0.47\textwidth}
    \includegraphics[width=7cm]{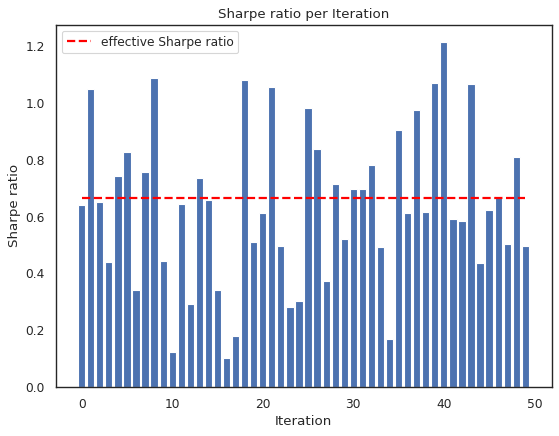}
    \caption{Out-of-sample Sharpe ratio per iteration.}
  \end{minipage}
\end{figure}

\newpage
\subsubsection{Feature set 6: Bitcoin + S\&P500 + Russell 2000 + EUR/USD + 10Y Treasury Yield + Gold/Silver}

\begin{figure}[h]
  \centering
  \begin{minipage}[b]{0.47\textwidth}
    \includegraphics[width=7cm]{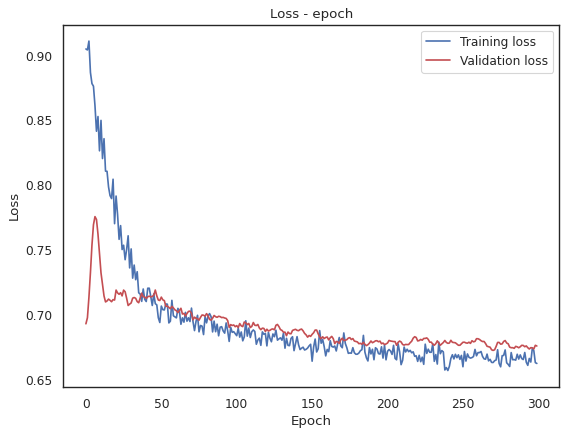}
    \caption{Train/Val. loss history of first iteration.}
  \end{minipage}
  \hfill
  \begin{minipage}[b]{0.47\textwidth}
    \includegraphics[width=7cm]{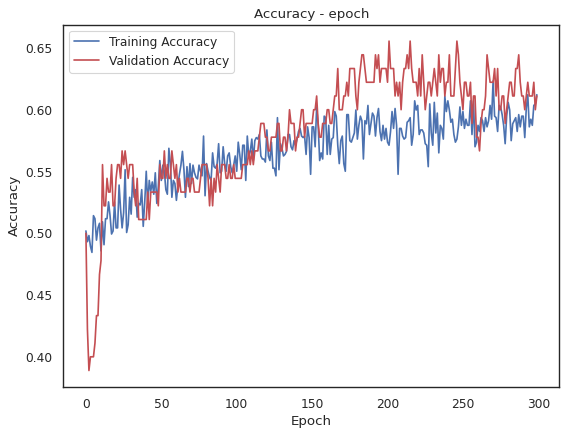}
    \caption{Train/Val. accuracy history of first iteration.}
  \end{minipage}
\end{figure}

\begin{figure}[h]
  \centering
  \begin{minipage}[b]{0.47\textwidth}
    \includegraphics[width=7cm]{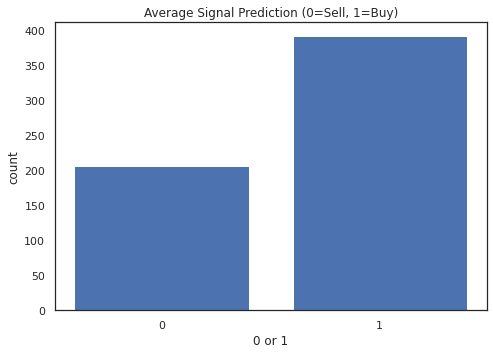}
    \caption{Average out-of-sample signals of 50 iterations.}
  \end{minipage}
  \hfill
  \begin{minipage}[b]{0.47\textwidth}
    \includegraphics[width=7cm]{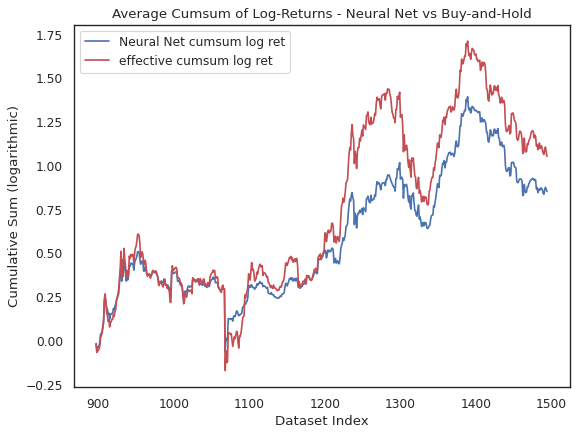}
    \caption{Average out-of-sample cumulative log returns of 50 iterations.}
  \end{minipage}
\end{figure}

\begin{figure}[h]
  \centering
  \begin{minipage}[b]{0.47\textwidth}
    \includegraphics[width=7cm]{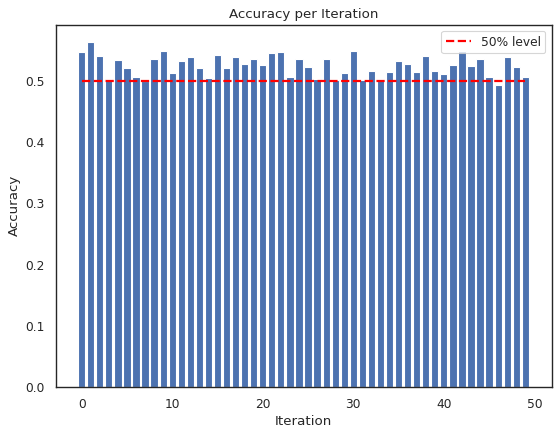}
    \caption{Out-of-sample accuracy per iteration.}
  \end{minipage}
  \hfill
  \begin{minipage}[b]{0.47\textwidth}
    \includegraphics[width=7cm]{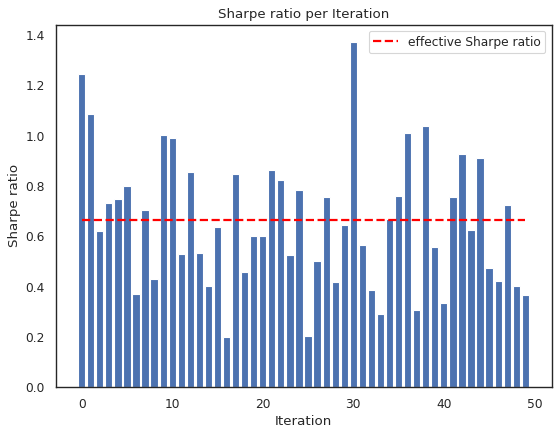}
    \caption{Out-of-sample Sharpe ratio per iteration.}
  \end{minipage}
\end{figure}

\end{document}